\documentclass[12pt]{article}

\usepackage{epsfig,latexsym,amsfonts,amsmath,amsthm,amssymb,amsbsy,multirow,slashed,wasysym,textcomp,dsfont,comment,mathtools,cancel,cite,diagbox,datetime,appendix,BOONDOX-calo}
\usepackage{tocloft}
\usepackage{bbold}
\usepackage{sidecap}
\usepackage{adjustbox}
\usepackage{pgfplots}
\usepackage{oplotsymbl}
\usepackage{tikzpagenodes}
\usepackage{adforn}
\usepackage{tikz}
\usetikzlibrary{backgrounds}
\usetikzlibrary{patterns}
\usetikzlibrary{arrows.meta}
\usetikzlibrary{shapes}
\tikzstyle{bag} = [align=center]
\usetikzlibrary{decorations.pathmorphing}
\usepackage[normalem]{ulem}
\usepackage[mathscr]{eucal}
\usepackage[version=3]{mhchem}
\usepackage{tikz-feynman}
\tikzfeynmanset{compat=1.0.0} 
\usetikzlibrary{shapes,arrows,positioning,automata,backgrounds,calc,patterns}

\tikzfeynmanset{
  fermion1/.style={
    /tikz/postaction={
      /tikz/decoration={
        markings,
        mark=at position 0.5 with {
          \node[
            transform shape,
            xshift=-0.5mm,
            fill,
            dart tail angle=100,
            inner sep=1.3pt,
            draw=none,
            dart
          ] { };
        },
      },
      /tikz/decorate=true,
    },
  },
  fermion2/.style={
    /tikz/postaction={
      /tikz/decoration={
        markings,
        mark=at position 0.5 with {
          \arrow{>[length=6pt, width=5pt]};
        },
      },
      /tikz/decorate=true,
    },
  },
}

\usepackage{hyperref}
\usepackage{subcaption}

 \newcommand{\badat}{\begin{alignedat}}
 \newcommand{\eadat}{\end{alignedat}}
 \newcommand\scalemath[2]{\scalebox{#1}{\mbox{\ensuremath{\displaystyle #2}}}}
 \def\be{\begin{equation}}
\def\ee{\end{equation}}

\def\p{\partial}

\usepackage{color}

\newcommand{\pink}[1]{\textcolor{\pink}{#1}}

\definecolor{dblue}{rgb}{0.2,0.50,0.80}

\def\n{k}

\def\O{\mathcal{O}}

\def\bh{{\bar h}}
\def\bz{{\bar z}}
\def\bw{{\bar w}}

\def\ba{{\bar a}}
\def\bb{{\bar b}}
\def\bc{{\bar c}}
\def\bd{{\bar d}}

\def\bh{{\bar h}}
\def\bz{{\bar z}}
\def\bw{{\bar w}}

\numberwithin{equation}{section} 
\pgfplotsset{compat=1.17} 
\begin{document}

 \begin{titlepage}
  \thispagestyle{empty}
  \begin{flushright}
  \end{flushright}
  \bigskip
  \begin{center}

                  \baselineskip=13pt {\LARGE \scshape{Lectures on Celestial Amplitudes}
         }

      \vskip1cm 

   \centerline{ 
   {Sabrina Pasterski}
   }

\bigskip\bigskip
 
 \centerline{ Princeton Center for Theoretical Science, Princeton, NJ 08544, USA}
 
\bigskip

\bigskip\bigskip

\end{center}

\begin{abstract}
  \noindent

Lecture notes prepared for the 2021 SAGEX PhD School in Amplitudes hosted by the University of Copenhagen August 10th through 13th.  Topics covered include: the manifestation of asymptotic symmetries via soft theorems, their organization into currents in a celestial CFT, aspects of the holographic dictionary, a literature guide, and accompanying exercises.

\end{abstract}

\end{titlepage}

\setcounter{tocdepth}{2}

\tableofcontents

\section{Introduction}

{\it Celestial Holography} proposes a duality between gravitational scattering in asymptotically flat spacetimes and a conformal field theory (CFT) living in two dimensions lower on the celestial sphere.  The central objects in this program are {\it Celestial Amplitudes}, which are $\cal{S}$-matrix elements written in a basis such that they transform as conformal correlators on this sphere.

A major theme that one encounters is the surprising extent to which infrared physics encodes infinite dimensional symmetry enhancements and points to a holographic description.  The aim of these lectures is to focus on the technology from scattering amplitudes that is used to motivate this duality, show how to translate this technology into the celestial basis, and then demonstrate some of the new techniques we gain by treating these amplitudes as correlators.

The organization of these lectures is as follows.  After discussing some of the motivations for this program, we start in section~\ref{scattering} by setting up the scattering problem from both a momentum space and a spacetime perspective.  This entails a review of the global symmetries, single particle states, and conformal compactification of Minkowski space.  In section~\ref{asymptotic} we explore how these global symmetries get enhanced, introducing the notion of asymptotic symmetries and how they manifest themselves as soft theorems in quantum field theory.  This leads us to define a holographic map from $\mathcal{S}$-matrix elements to conformal correlators living on the celestial sphere in section~\ref{dictionary}.  After defining the map for single particle states we dive deeper into how various structures of amplitudes are encoded in this basis.  We close with a review of recent efforts as well as a guide to the earlier literature in section~\ref{literature}.  Relevant exercises accompany each section.

\subsubsection*{A new framework for scattering}
The celestial holography program is an attempt to apply the holographic principle to spacetimes with vanishing cosmological constant.  On the one hand this is motivated by the success of AdS/CFT~\cite{Maldacena:1997re,Witten:1998qj, Aharony:1999ti}.  Indeed, the ability to similarly apply such a duality to answer non-perturbative questions about quantum gravity, as well as have a top-down stringy construction of celestial CFTs are ultimate targets.  However, it can also be viewed as a CFT-inspired take on the amplitudes and earlier $\mathcal{S}$-matrix programs~\cite{Eden:1966dnq,Elvang:2013cua,Henn:2014yza}:  if we have an intrinsic description of consistent celestial CFTs, we can hope to use CFT-bootstrap style machinery to compute and constrain scattering. 

The central object of study is the $\mathcal{S}$-matrix.  The advantage we have is that we are building, from the bottom up, a description in which the symmetries are not only front and center, but also infinite dimensional.  The impetus for this sub-field was the keen observation by Strominger~\cite{Strominger:2013jfa,Strominger:2013lka} that soft theorems in quantum field theory~\cite{Weinberg:1965nx} are manifestations of Ward identities for asymptotic symmetries~\cite{Bondi:1962px,Sachs:1962wk,Sachs:1962zza}.  Even more beautifully, the generators for these asymptotic symmetries looked like currents living two dimensions lower, on the celestial sphere.  The subleading soft graviton theorem~\cite{Cachazo:2014fwa} corresponding to an asymptotic Virasoro symmetry~\cite{Kapec:2014opa} even gave a candidate stress tensor~\cite{Kapec:2016jld}. However, for the conformal Ward identity to take the standard form, one needs to change from momentum to boost eigenstates~\cite{Pasterski:2016qvg,Pasterski:2017kqt,Pasterski:2017ylz}.  This story will be the focus of section~\ref{asymptotic} and guide us to our holographic map in section~\ref{dictionary}.

Now the $\mathcal{S}$-matrix is already holographic via the manner in which we can use an object that depends on-shell data to learn about bulk physics.  The dictionary between external single particle states and (quasi)-primary operators in celestial CFT is simply an integral transform on the external on-shell states that implements our change of basis
\be\label{map}
~_{\mathrm{boost}}\langle out |\mathcal{S}|in\rangle_{\mathrm{boost}}=\langle \mathcal{O}^{\pm}_{\Delta_1,J_1}(z_1,\bz_1)...\mathcal{O}^{\pm}_{\Delta_1,J_1}(z_n,\bz_n)\rangle_{\mathrm{CCFT}}
\ee
where the $\pm$ superscripts label operators corresponding to incoming  and outgoing states.   This recasting has interesting implications. In section~\ref{dictionary} we will focus on how it resolves soft limits at different orders in $\omega\rightarrow 0$ to poles at special values of the conformal dimension where celestial currents appear. We will also see that collinear limits play a fundamental role in this setup, giving the OPE data for CCFT which, in turn, is highly constrained by the symmetries.  Furthermore, celestial amplitudes probe scattering at all energy scales -- a fact which inverts the standard Wilsonian paradigm and gives an interesting route to try to constrain the UV.

We will use our closing section to survey some of the active research lines within celestial holography and show how the tools we are using and developing connect to adjacent sub-fields. Before commencing our exploration of celestial amplitudes, we should point out other helpful pedagogical references on this topic. For recent reviews on how infrared physics leads to celestial holography see~\cite{Strominger:2017zoo,Pasterski:2019ceq,Raclariu:2021zjz}.  The reviews~\cite{Elvang:2013cua,Oblak:2015qia,Banados:2016zim,Taylor:2017sph,Compere:2018aar} offer more in-depth coverage of some of the machinery on symmetries, general relativity, and amplitudes that we employ herein.

\section{Kinematics of Scattering}\label{scattering}

Throughout these lectures we will be considering 4D Minkowski space, for which we have a 2D celestial sphere.  We will start by examining the scattering problem from both a spacetime and momentum space perspective, focusing on the global spacetime symmetries and how they act on the celestial sphere and the external scattering data.  This requires reviewing the causal structure of Minkowski space.  We will also use this section to set up various notation and coordinate conventions.

\subsection{Global Symmetries}

Lorentz transformations in $\mathbb{R}^{1,d+1}$ act as conformal transformations in $\mathbb{R}^{d}$.  The aim of this section is to review these global symmetries in a manner that will prepare us to generalize to asymptotic symmetries in the following section, and set up notation we will be using hereafter.  The review~\cite{Oblak:2015qia} is a nice resource dedicated to this topic.

\subsubsection*{Isometries of Minkowski Space}\label{isom}
We begin by reviewing the isometries of 3+1 dimensional Minkowski space, starting with the line element in Cartesian coordinates
\be\label{etaflat}
ds^2=\eta_{\mu\nu}dx^\mu dx^\nu=-(dX^0)^2+(dX^1)^2+(dX^2)^2+(dX^3)^2.
\ee
The infinitesimal coordinate transformations $X^\mu\mapsto X^\mu+\xi^\mu$ that preserve this metric
\be\label{lieder}
\mathcal{L}_{\xi} \eta_{\mu\nu}=\p_\mu \xi_\nu +\p_\nu \xi_\mu=0
\ee
are parameterized by a constant four vector $b^\mu$ and antisymmetric tensor $\omega_{\mu\nu}$
\be
\xi^\mu =\omega^\mu_{~\nu} X^\nu+b^\mu,~~~\omega_{\mu\nu}=-\omega_{\nu\mu},
\ee
which correspond to infinitesimal translations and Lorentz transformations.  Upon taking Lie brackets of the corresponding vector fields
\be\label{poincare}
P_\mu=-i\p_\mu,~~~M_{\mu\nu}=i(x_\mu\p_\nu-x_\nu\p_\mu),
\ee
we see that these generators  obey the Poincar\'e algebra
\be\badat{3}
[P_\mu,P_\nu]&=0,~~~[P_\rho,M_{\mu\nu}]=i(\eta_{\mu\rho}P_\nu-\eta_{\nu\rho}P_\mu),\\
[M_{\mu\nu},M_{\rho\sigma}]=&~i(\eta_{\mu\rho}M_{\nu\sigma}-\eta_{\mu\sigma}M_{\nu\rho}-\eta_{\nu\rho}M_{\mu\sigma}+\eta_{\nu\sigma}M_{\mu\rho}).
\eadat\ee
Now the finite versions of these transformations are elements of the Poincar\'e group $O(1,3)\ltimes\mathbb{R}^{1,3}$ which takes the form of a semi-direct product of Lorentz transformations 
\be
O(1,3)=\{\Lambda\in GL(4,\mathbb{R})|\Lambda^\top \eta\Lambda=\eta\}
\ee
 and translations. The Lorentz group is the subgroup of Poincar\'e which stabilizes the spacetime origin. It has four disconnected components related by parity and time reversal to the proper orthochronous Lorentz group
\be
SO^+(1,3)=\{\Lambda\in O(1,3)|\det(\Lambda)=+1, \Lambda^0_{~0}\ge1\}
\ee
that one reaches by exponentiating the infinitesimal transformations above.

The following isomorphism will be central to what follows
\be\label{so31sl2c}
SO^+(1,3)\cong SL(2,\mathbb{C})/\mathbb{Z}_2.
\ee
Writing  a generic $M\in SL(2,\mathbb{C})$ as
\be\label{sl2c}
M=\left(\begin{array}{cc}
    a & -c \\
    -b & d
\end{array}
\right),~~~ad-bc=1
\ee
for $\{a,b,c,d\}\in \mathbb{C}$, we can demonstrate this isomorphism as follows.  Using the Pauli matrices we can define
\be
(\sigma^\mu)_{a\dot{b}}=(\mathbb{1},\sigma^i)_{a\dot{b}},~~~(\bar{\sigma}^\mu)^{\dot{a}b}=(\mathbb{1},-\sigma^i)^{\dot{a}b}
\ee
which let us map $4$-vectors to $2\times2$ matrices and back
\be\label{isom}
{v}_{a\dot{b}}=v_\mu(\sigma^\mu)_{a\dot{b}} ~~~v^\mu=-\frac{1}{2}\mathrm{tr}(v\bar{\sigma}^\mu).
\ee
The intertwining properties of the Infeld-van der Waerden symbols imply that for $M$ in~\eqref{sl2c} we have
\be
M^\dagger X^\mu\sigma_\mu M=(\Lambda^\mu_{~\nu}X^\nu)\sigma_\mu
\ee
where 
\begin{equation}\label{Lambda}
   \scalemath{0.9}{ \Lambda^\mu_{~\nu}=
   \frac{1}{2}
    \scalemath{.95}{
    \left(\begin{array}{cccc}
    a\ba+b\bb+c\bc+d\bd & a\bb+\ba b+\bc d+c\bd & i(a\bb-\ba b+c \bd-\bc d) & b\bb-a\ba-c\bc+d\bd\\
    a\bc+\ba c+b\bd+\bb d & a \bd+\ba d +b\bc+\bb c & i(a\bd -\ba d -b\bc+\bb c) & b\bd+\bb d-a\bc -\ba c\\
    i(\ba c-a\bc -b \bd +d\bb ) & i(\ba d-a\bd -b \bc+\bb c) & a\bd+\ba d-b\bc -\bb c & i(a\bc -\ba c-b \bd+\bb d)\\
     c\bc +d\bd-a \ba- b\bb  & c\bd+\bc d -a \bb- \ba b& i(\ba b-a\bb+c\bd -\bc d) & a\ba -b\bb -c\bc +d\bd
    \end{array}\right)\,
    }}.
\end{equation}
It is straightforward to verify that this satisfies $\det \Lambda=1$ and $\Lambda^0_{~0}\ge1$.  Note also an invariance under the reflection $M\mapsto -M$. This is the $\mathbb{Z}_2$ quotient in~\eqref{so31sl2c}. 

\subsubsection*{Conformal Isometries of the Riemann Sphere}
Now $SL(2,\mathbb{C})/\mathbb{Z}_2$ is also the group of global conformal transformations of the Riemann sphere.  Recall that conformal transformations preserve angles but not distances
\be
g'_{AB}(x')=\Omega^2(x)g_{AB}(x).
\ee
Here we will start with the flat Euclidean metric
\be\label{2dplane}
ds^2
=\delta_{AB}dx^A dx^B.
\ee
We now want to demand
\be\label{conf}
\mathcal{L}_{\xi} \delta_{AB}=\p_A \xi_B +\p_B \xi_A=2\omega(x)\delta_{AB}
\ee
where $\Omega=1+\omega$, and upon taking a trace we see that $\omega=\p_A\xi^A$. 

Let us first look at the solutions one would get if we let $A=\{1,....d\}$ for generic $d$.  We have
\be\badat{3}\label{2con}
P_A=-i\p_A,~~~~&J_{AB}=i(x_A\p_B-x_B\p_A)\\
D_A=-ix^A\p_A,~~~~&K_{A}=i(x^2\p_A-2x_Ax^B\p_B)
\eadat\ee
which obey the following algebra
\begin{gather}
[D,K_A]=-iK_A,~~~[D,P_A]=iP_A,~~~
[K_A,P_B]=2i(\delta_{AB}D-J_{AB})\notag\\
[K_A,J_{BC}]=i(\delta_{AB}K_C-\delta_{AC}K_B),~~~[P_A,J_{BC}]=i(\delta_{AB}P_C-\delta_{AC}P_B)\\
[J_{AB},J_{CD}]=i(\delta_{AC}J_{BD}-\delta_{AD}J_{BC}-\delta_{BC}J_{AD}+\delta_{BD}J_{AC})\notag
\end{gather}
with all other Lie brackets vanishing.  We see that, in addition to translations and rotations, we have dilatations which generate scale transformations and special conformal transformations whose exponentiated version implements a translation on the inverted coordinate
\be
\frac{x^A}{x^2}\mapsto \frac{x^A}{x^2}+b^A.
\ee
This gives a total of $\frac{1}{2}(d+2)(d+1)$ symmetries matching the number of generators of $SO(1,d+1)$.  Indeed, we can identify the Lorentz generators in $\mathbb{R}^{1,d+1}$ with the conformal generators in $\mathbb{R}^{d}$ as follows
\be
D=M_{d+1,0},~~~P_A=M_{A d+1}-M_{A0},~~K_A=M_{A, d+1}+M_{A0},~~~
J_{AB}=M_{AB}.
\ee
In many contexts, it is natural to think of $\mathbb{R}^{1,d+1}$ as an embedding space. Lorentz transformations map the lightcone of the origin in $\mathbb{R}^{1,d+1}$ to itself.  If we embed $\mathbb{R}^d$ into this lightcone via the null vector
\be
q^\mu(x)=(1+x^2,x^A,1-x^2)
\ee
then the conformal transformation $x\mapsto x'$ corresponds to a boost
\be\label{emb}
q^\mu(x')=\frac{(\Lambda q)^\mu}{(\Lambda q)^+},
\ee
up to rescaling the Lorentz transformed $q^\mu$ back to the canonical section $q^+\equiv\frac{1}{2}(q^0+q^3)=1$.
The embedding space is convenient from the practical point of view that the Lorentz transformations act linearly on $\mathbb{R}^{1,d+1}$, in contrast to the corresponding transformation of the $x^A$. If we specify to $d=2$, it is natural to introduce  the complex coordinates
\be\label{wbw}
w=x^1+ix^2,~~~\bw=x^1-ix^2.
\ee
The reference direction embedding the complex $w$ plane into the lightcone of  $\mathbb{R}^{1,3}$ takes the form
\be\label{qref}
q^\mu(w,\bw)=(1+w\bw,w+\bw,i(\bw-w),1-w\bw).
\ee
The finite conformal transformations are the M\"obius transformations
\be\label{mobius}
w\mapsto w'=\frac{aw+b}{cw+d},~~~\bw\mapsto w'=\frac{\bar{a}\bw+\bar{b}}{\bar{c}\bw+\bar{d}}
\ee
parameterized by $M\in SL(2,\mathbb{C})/\mathbb{Z}_2$ as in~\eqref{sl2c} above, and where the $\Lambda$ appearing in the $d=2$ analog of~\eqref{emb} is given by~\eqref{Lambda}.  
We see that the point $w=-d/c$ gets mapped to infinity.  Adding this point takes us to the conformal compactification $\hat{\mathbb{C}}$, the Riemann sphere, on which the M\"obius transformations act bijectively.

However in $d=2$ there is actually a richer set of infinitesimal transformations that locally preserve the conformal class of metrics.
In the coordinates~\eqref{wbw} the metric~\eqref{2dplane} takes the form
\be
ds^2
=dwd\bw
\ee
and equation~\eqref{conf} is particularly simple to solve. We find
\be
\xi^w=Y(w),~~~\xi^\bw=\bar{Y}(\bw)
\ee
where we demand $\bar{Y}(\bw)=Y(w)^*$ for real diffeomorphisms of the $(x^1,x^2)$ plane.   The corresponding Laurent modes 
\be
L_n=-w^{n+1}\p_w,~~~{\bar L}_n=-\bar{w}^{n+1}\p_{\bar w},
\ee
for $n\in \mathbb{Z}$, obey two copies of the Witt algebra
\be\label{lm}
[L_m,L_n]=(m-n)L_{m+n},~~~[\bar{L}_m,\bar{L}_n]=(m-n)\bar{L}_{m+n},~~~[L_n,\bar{L}_m]=0
\ee
where the $\mathfrak{sl}(2,\mathbb{C})$ subalgebra is spanned by $n=\{-1,0,1\}$.   In terms of the standard presentation of the Lorentz generators we have
\begin{equation}
\label{embedding_generators}\scalemath{0.95}{
    \badat{3}
 L_0&=\frac{1}{2}(J_3-i K_3) \, ,~~ & L_{-1}&=\frac{1}{2}(-J_1+i J_2+i K_1+K_2)
  \, ,~~
    & L_1&=\frac{1}{2} (J_1+i
   J_2-i K_1+K_2)
    \, ,~~
   \\
 \bar L_0&=\frac{1}{2} (-J_3-i K_3)
  \, ,~~&
   \bar L_{-1}&=\frac{1}{2} (J_1+i J_2+i K_1-K_2)
    \, ,~~
   & \bar L_1 &=\frac{1}{2} (-J_1+i J_2-i K_1-K_2)
    \, ,~~
   \\
\eadat}
\end{equation}
where we have introduced notation for the boosts and rotation generators
\be
K_i=M_{i0},~~~J_i=\frac{1}{2}\epsilon_{ijk}M^{jk}
\ee
which we will use in the following section. From the way that we've embedded the Riemann sphere into the lightcone of Minkowski space, one might not expect these extra generators to be considered symmetries of our 4D spacetime.  However in section~\ref{asymptotic}, we will see that once we include gravity there is a larger class of spacetime symmetries to consider called asymptotic symmetries.  We will furthermore be able to define currents that obey these infinite dimensional symmetry algebras using the low energy dynamics of scattering.

\subsection{Scattering States in Momentum Space}\label{momentum}

In this section we review how the global symmetries are used to classify the single particle external states appearing as external states in the momentum space $\mathcal{S}$-matrix. A good reference for this section is~\cite{Weinberg:1995mt}. The CCFT generalization of these methods, which we will return to later, is covered in~\cite{Banerjee:2018gce}. We will also use this as an opportunity to set up our spinor helicity conventions, which follow~\cite{Elvang:2013cua}.

\subsubsection*{Single Particle States}

Single particle states are irreducible representations of the Poincar\'e group.  The Poincar\'e isometries should be unitarily realized on our 4D Hilbert space.\footnote{The fact that the Poincar\'e group is realized unitarily on the Hilbert space implies the generators  $\{P_\mu, J_i,K_i\}$ are Hermitian, which we note implies
\be
L_{i}^\dagger=-\bar{L}_i,~~~\bar{L}_{i}^\dagger=-L_i.
\ee}
If we start with a state with four momentum $p^\mu$ then we want
\be\label{induced}
U(\Lambda)|p\rangle=|\Lambda p\rangle.
\ee
In the standard construction, the single particle states are built as induced representations of the little group which stabilizes a fixed reference on-shell momentum. Letting
\be\label{masslessref}  
p_{ref}^\mu=(m,0,0,0),~~~p_{ref}^\mu=(\omega,0,0,\omega)=\omega q^\mu(0,0)
\ee
for massive and massless fields, respectively, we can get to any other on-shell momentum by an appropriate boost
\be
p^\mu=L^\mu_{~\nu}(p)p_{ref}^\nu.
\ee
Defining the state 
\be
|p\rangle=U(L(p))|p_{ref}\rangle,
\ee
 we see that
\be
U(\Lambda)|p\rangle=U(\Lambda)U(L(p))|p_{ref}\rangle=U(L(\Lambda p))U(L^{-1}(\Lambda p)\Lambda L(p))|p_{ref}\rangle
\ee
where $L^{-1}(\Lambda p)\Lambda L(p)$ is in the little group stabilizing $p_{ref}^\mu$, while $U(L(\Lambda p))$ takes us from $|p_{ref}\rangle$ to $|\Lambda p\rangle$, reproducing the right side of~\eqref{induced}. We thus see how to build representations of the Lorentz group from representations of the little group.  

The quadratic Casimirs of the Poincar\'e group are the squares of the four momentum $P_\mu$ and the Pauli-Lubanski pseudo-vector
\be\label{plvec}
W_\mu=\frac{1}{2}\epsilon_{\mu\rho\sigma\lambda}M^{\rho\sigma}P^\lambda.
\ee
Since these operators commute with all of the Poincar\'e generators, their eigenvalues will label our irreducible representations.  For massive fields the little group is just the rotations generated by the $J_i$.   The quadratic Casimirs evaluate to
\be
P^2=-m^2,~~~W^2=-m^2s(s+1)
\ee
for the spin-$s$ representation of $SO(3)$.  Meanwhile, the $m=0$ little group is generated by
\be
\{J_3,J_1+K_2,J_2-K_1\}\Leftrightarrow \{L_0-\bar{L}_0,L_1,\bar{L}_1\}
\ee
whose algebra is isomorphic to the symmetries of the 2-plane $ISO(2)$. The quadratic Casimirs are now
\be\label{w2}
P^2=0,~~~W^2=\omega^2((J_1+K_2)^2+(J_2-K_1)^2)=\omega^2L_{1}\bar{L}_{1}.
\ee
Excluding continuous spin representations amounts to demanding that the states transform trivially under the translation generators of $ISO(2)$.  These have $W^2=0$ and are labeled by the $U(1)$ rotation eigenvalue that corresponds to the helicity $\ell$.

In what follows we will label the single particle states via coordinates suited to their on-shell momenta.  For instance for $p^2=-m^2$ we use 
\be\label{massive}
|y,z,\bz\rangle:=|p\rangle  ,~~~~p^\mu=\frac{m}{2y}(1+y^2+z\bz,z+\bz,i(\bz-z),1-y^2-z\bz)
\ee
where $y,z,\bz$ are coordinates on the unit hyperboloid.  For $p^2=0$ we will write
\be\label{hb}
|\omega,w,\bw\rangle:=|p\rangle,~~~~p^\mu=\omega q^\mu(w,\bw)
\ee
suppressing the spin/helicity and mass labels.
We will mainly focus on the massless case in what follows. In terms of this notation, and restoring the helicity index, we have
\be
L_1|\omega,0,0;\ell\rangle=\bar{L}_1|\omega,0,0;\ell\rangle=0,~~~(L_0-\bar{L}_0)|\omega,0,0;\ell\rangle=\ell|\omega,0,0;\ell\rangle.
\ee

\subsubsection*{Free Field Expansions}

In the free theory one can build multi-particle states from tensor products of these single particle states.  The Fock space is then constructed as a direct sum of these $n$-particle sectors. If one assumes the interactions die off fast enough with separation, then the asymptotic initial and final states should be described by those of the free theory.\footnote{Long range interactions that spoil this are tied to the IR divergences we will study in the following section.}  The overlap between preparing a fixed multi-particle $in$-state and it evolving to a fixed multi-particle $out$-state is given by the $\mathcal{S}$-matrix
\be
\langle out|\mathcal{S}|in\rangle,~~~
|in\rangle=a_{p_1}^\dagger...a_{p_m}^\dagger|0\rangle,~~~|out\rangle=a_{p_{m+1}}^\dagger...a_{p_n}^\dagger|0\rangle
\ee
which will be our central object of study.  In quantum field theory, $\mathcal{S}$-matrix elements are typically computed using the LSZ formalism starting from Fourier transforms of time-ordered correlators and taking residues as the momenta go on shell. Its derivation starts by extracting the creation and annihilation operators using an inner product of the free plane wave solutions and the field operator on early and late time slices. 

This starting point will be used to generalize to other wavepackets for the external scattering states in what follows.  We will go through the ingredients needed for the massless spin-1 and spin-2 cases here. The Maxwell and Einstein-Hilbert actions are given by
\be
S_{EM}=-\frac{1}{4e^2}\int d^4x \sqrt{-g}F_{\mu\nu}F^{\mu\nu},~~~{S}_{EH}=\frac{1}{4\kappa}\int d^4x \sqrt{-g}R
\ee
in terms of the field strength $F_{\mu\nu}=\p_\mu A_\nu-\p_\nu A_\mu$ and Ricci scalar.  Here $\kappa=\sqrt{32\pi G}$. 
  The linearized source-free equations of motion take the form
\be\label{aheom}
\Box A_\mu-\p_\mu\p^\nu A_\nu=0,~~~~\partial_\sigma \partial_\nu h^\sigma_{~\mu}+\partial_\sigma \partial_\mu h^\sigma_{~\nu}-\partial_\mu \partial_\nu h-\Box h_{\mu\nu}=0
\ee
where we have expanded $g_{\mu\nu}=\eta_{\mu\nu}+h_{\mu\nu}$.  The corresponding field operators then have the following free mode expansions 
\begin{equation}\label{modexp2}
\badat{3}
    \hat{A}_\mu(X)&=e\sum_{\alpha\in\pm}\int \frac{d^3k}{(2\pi)^3}\frac{1}{2k^0}\left[\epsilon_\mu^{\alpha *}a_\alpha e^{ik\cdot X}+\epsilon_\mu^{\alpha}a_\alpha^\dagger e^{-ik\cdot X}\right]\,,\\
    \hat{h}_{\mu\nu}(X)&=\kappa\sum_{\alpha\in\pm}\int \frac{d^3k}{(2\pi)^3}\frac{1}{2k^0}\left[\epsilon_{\mu\nu}^{\alpha *}a_\alpha e^{i k\cdot X}+\epsilon_{\mu\nu}^{\alpha}a_{\alpha}^\dagger e^{-ik\cdot X}\right]\,,
    \eadat
\end{equation}
where for momentum $k^\mu=\omega q^\mu$ we use the polarization vectors
\be\label{polar}
\epsilon_+^\mu=\frac{1}{\sqrt{2}}\p_w q^\mu,~~~\epsilon_-^\mu=\frac{1}{\sqrt{2}}\p_\bw q^\mu
\ee
and $\epsilon_\pm^{\mu\nu}=\epsilon_\pm^\mu\epsilon_\pm^\nu$. Since spinor helicity variables often appear in the celestial amplitudes literature, let us briefly comment on our conventions.  Using~\eqref{isom} to go from four-vectors to bi-spinors, $k=\pm \omega q^\mu$ maps to
\be\label{spinorhelicity}
k_{a\dot{a}}=-|k]_a\langle k|_{\dot{a}},~~~
|k]_a=\sqrt{2\omega}\left(\begin{array}{c}
     \bw \\
     -1 
\end{array}\right)_a,~~~\langle k|_{\dot a}=\pm\sqrt{2\omega}\left(\begin{array}{c}
     w \\
     -1 
\end{array}\right)_{\dot a}
\ee
where indices are raised and lowered with the Levi-Civita tensor. The polarization vectors~\eqref{polar} match the more general
\be
\epsilon_+=-\frac{\langle r|\bar{\sigma}^\mu|k]}{\sqrt{2}\langle r k\rangle},~~~\epsilon_-=-\frac{\langle k|\bar{\sigma}^\mu|r]}{\sqrt{2}[rk]}
\ee
for the choice of reference spinor $|r]_a=(1,0)$. Our choice of normalization is such that the mode operators obey the following commutation relations
\be
[a_{\alpha}(\vec{p}),a^\dagger_{\alpha'}(\vec{p}')]=2\omega_p(2\pi)^3\delta_{\alpha,\alpha'}\delta^{(3)}(\vec{p}-\vec{p'}).
\ee These creation/annihilation operators can be extracted from the mode expansions by taking an inner product with the corresponding plane wave solution.  For example
\be
a_\pm(p)=i(\hat{A},(\epsilon^{\pm}e^{ip\cdot X})^*)
\ee
and similarly for spin-2, where the inner product is given by
\be\badat{3}
(A,A')&=-i\int d\Sigma^\mu [A^\nu( \nabla_\mu A'^*_\nu- \nabla_\nu A'^*_\mu)-(A\leftrightarrow A'^*)],\\
(h,h')&=-i\int d\Sigma^\rho [h^{\mu\nu}( \nabla_\rho h'^*_{\mu\nu}- 2\nabla_\mu h'^*_{\rho\nu})-(h\leftrightarrow h'^*)],
\eadat\ee
restricting to trace-free spin-2 perturbations. We will be applying the same procedure with modified wavepackets when we set up our holographic dictionary in section~\ref{dictionary}.

This brings us to a comment on the $SL(2,\mathbb{C})$ representations of these local operators.  Because the Lorentz group is non-compact, there are no faithful finite dimensional unitary representations. This does not prevent us from the single particle states forming a unitary representation since those are infinite dimensional representations induced from finite dimensional representations of the little group.   However, the vector indices on our field operators do transform under finite dimensional, non-unitary representations. The reconciliation of how both these representations can appear in our operator mode expansion boils down to gauge invariance.  The $ISO(2)$ translations shift the polarization vector parallel to $k^\mu$, while the single particle states transformed trivially under these generators.  This amounts to 
\be\label{eq:gaugetransf}
U(\Lambda)\hat{A}_\mu(x) U^{-1}(\Lambda)=\Lambda^\nu_{~\mu}\hat{A}_\nu(\Lambda x)+\p_\mu\hat{\Omega}(x,\Lambda).
\ee
In the end, our $\mathcal{S}$-matrix elements are gauge invariant, so that everything is consistent.  We will take advantage of this when we construct convenient gauge representatives of the conformal primaries in section~\ref{dictionary}.  Curiously the gauge transformations~\eqref{eq:gaugetransf} are indeed `large' in the asymptotic symmetry sense~\cite{Hamada:2017uot} that we will study in the next section.

\subsection{Data at the Conformal Boundary}\label{nullinfty}

When setting up the scattering problem in position space, we want to specify field configurations on early and late time Cauchy surfaces.  An important step is to understand the causal structure of Minkowski space.   We will see that there are different notions of infinity.  In particular, the conformal boundary includes null components where we will specify the free data for massless excitations.

\begin{figure}[th!]
\centering
\begin{tikzpicture}[scale=2.3]
\newcommand{\cross}{$\mathbin{\tikz [x=.75ex,y=.75ex,line width=.25ex, blue] \draw (0,0) -- (1,1) (0,1) -- (1,0);}$}
\definecolor{darkgreen}{rgb}{.0, 0.5, .1};
\draw[thick](0,0) --(1,1) node[right] {$i^0$} --(0,2)node[above] {$i^+$} --(-1,1) --(0,0)  node[below] {$i^-$} ;
\draw[red,<->,thick](1+.03,1+.03) -- node[above]{$~u$} (0+.03,2+.03);
\draw[darkgreen,<->,thick](1+.03,1-.03) -- node[below]{$~v$} (0+.03,0-.03);
\draw[thick] (-1,1.8) circle (.6em);
\draw[blue,fill=blue] (-1,2.04) circle (.75pt);
\draw[blue,fill=blue] (.5,1.5) circle (.75pt);
\draw[blue,fill=blue] (-.5,.5) circle (.75pt);
\node at (-1,1.555) {\cross};
\node at (-.5,1.5) {\cross};
\node at (.5,.5) {\cross};
\node[blue] at (-1, 2.15) {$z$};
\node at (1/2+.3,3/2+.3) {$\cal{I}^+$};
\node at (1/2+.3,1/2-.3) {$\cal{I}^-$};
\end{tikzpicture}
\caption{Penrose diagram for Minkowski space.  Massive particles enter and exit at past ($i^-$) and future ($i^+$) timelike infinity.  Massless excitations enter and exit at past ($\cal{I}^-$) and future ($\cal{I}^+$) null infinity.  Meanwhile, the boundary of a Cauchy slice will limit to spatial infinity ($i^0$).
\label{penrose}}
\end{figure}
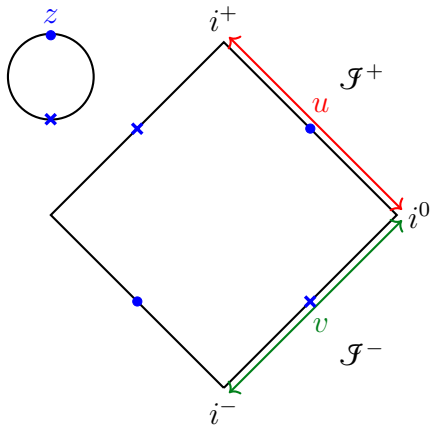

We can see this as follows.  Starting from the retarded and advanced time coordinates
\be
u=t-r,~~~v=t+r
\ee
we can introducing rescaled coordinates
\be
u=\tan U,~~~v=\tan V
\ee
so that the spacetime is spanned by $U,V\in(-\frac{\pi}{2},\frac{\pi}{2})$. The metric in the new timelike and radial coordinates
\be
T=U+V,~~~R=V-U
\ee
is conformally related to a patch of $S^3\times\mathbb{R}$
\be
ds^2=\Omega^{-2}(-dT^2+dR^2+2\sin^2R \gamma_{z\bz}dzd\bz),~~~~\Omega^{2}=4\cos^2U \cos^2 V.
\ee
The conformally rescaled metric $\Omega^2ds^2$ captures the causal structure of the original spacetime.  The following points in the $(R,T)$ plane play a special role
\begin{itemize}
    \item Massive particles follow trajectories that begin at {\it past timelike infinity} $i^-$ at $(R,T)=(0,- \pi)$, and end at  {\it future timelike infinity} $i^+$ at $(R,T)=(0,+\pi)$. 
       \item Massless trajectories begin at {\it past null infinity} $\mathcal{I}^-$ parameterized by $U=-\frac{\pi}{2},~~V\in(-\frac{\pi}{2},\frac{\pi}{2})$, and end at  {\it future null infinity} $\mathcal{I}^+$ parameterized by $V=+\frac{\pi}{2},~~U\in(-\frac{\pi}{2},\frac{\pi}{2})$.
       \item Spacelike geodesics limit to {\it spatial infinity} $i^0$ at $(R,T)=(\pi,0)$.
\end{itemize}
The Penrose diagram for Minkowski space is illustrated in figure~\ref{penrose}.  Massless trajectories move on $45^\circ$ diagonals. Typically one would draw only the right half of the figure and there would then be an $S^2$ over each point in the $(r,t)$ plane aside from the locus $r=0$. Here we've shown both the north and south poles of this transverse $S^2$ since the antipodal matching condition across spatial infinity will play an important role in what follows.

\pagebreak

\vspace{1em}

The conformal compactification will inform our coordinate choices in what follows.  To capture outgoing radiative data near future null infinity, we use retarded radial coordinates $(u,r,z,\bz)$
\be\label{ubondi}
X^\mu=\left(u+r,r\frac{z+\bz}{1+z\bz},ir\frac{\bz-z}{1+z\bz},r\frac{1-z\bz}{1+z\bz}\right)
\ee
for which the flat metric~\eqref{etaflat} takes the form
\be\label{ubondi2}
ds^2
=-du^2-2dudr+2r^2\gamma_{z\bz}dzd\bz,~~~~~\gamma_{z\bz}=\frac{2}{(1+z\bz)^2},
\ee
where $\gamma_{z\bz}$ is the round sphere metric in stereographic coordinates for the Riemann sphere $z=e^{i\phi}\tan\frac{\theta}{2}$.  Meanwhile near past null infinity we use the advanced time coordinate $v=t+r$ and an antipodally identified $S^2$ coordinate $z\mapsto-\frac{1}{\bz}$, for reasons  we will motivate in the following section.

This $S^2$ cross section of null infinity is the celestial sphere. In the previous section we saw that single particle states for massless particles are specified by a direction and an energy.  We will now see how this maps to the conformal boundary. Starting from the mode expansions~\eqref{modexp2}, we can approach future null infinity by holding $u$ fixed as we take $r$ to infinity.  The rapidly oscillating phase localizes the direction in momentum to be parallel to the position space direction towards which the particle scatters
\be
\lim\limits_{r\rightarrow\infty}\sin\theta e^{i\omega q^0r(1-\cos\theta)}=\frac{i}{\omega q^0 r}\delta(\theta)+\mathcal{O}((\omega q^0 r)^{-2}),
\ee
so that the field operators in this limit take the form
 \begin{equation}\label{eq:hatA}
    \lim\limits_{r\rightarrow \infty} \hat{A}_z=-\frac{ie}{\sqrt{2}(2\pi)^2} \int_0^\infty d\omega \left[a_+(\omega,z,\bz)e^{-i\omega (1+z\bz) u}-a^\dagger_-(\omega,z,\bz)e^{i\omega (1+z\bz) u}\right]\,,
 \end{equation}
 and
 \begin{equation}\label{eq:hatH}
  \lim\limits_{r\rightarrow \infty}\frac{1}{r}\hat{h}_{zz}=-\frac{i\kappa}{(2\pi)^2} \frac{1}{(1+z\bz)} \int_0^\infty d\omega \left[a_+(\omega,z,\bz)e^{-i\omega (1+z\bz) u}-a^\dagger_-(\omega,z,\bz)e^{i\omega (1+z\bz) u}\right]\,.
 \end{equation}
We thus see that we can create a state of definite energy by Fourier transforming along the $u$ direction
\be\label{pout}
\langle p,+1|=\lim\limits_{r\rightarrow\infty}\int du e^{i\omega(1+z\bz) u}\langle 0|\hat{A}_z,~~~\langle p,+2|=\lim\limits_{r\rightarrow\infty}\frac{1}{r}\int du e^{i\omega(1+z\bz) u}\langle 0|\hat{h}_{zz}
\ee
where $p_\mu=\omega q_\mu(z,\bz)$.  For massive fields there is a similar saddle point one can take~\cite{Campiglia:2015qka}.  If we resolve timelike infinity with hyperboloids of constant $X^2$, then massive particles with momenta $p^2=m^2$ will asymptote to trajectories that where $\hat{p}\propto \hat{x}$.

\vspace{1em}

\noindent{\bf Exercise:} 
This exercise will use the machinery we've reviewed here to anticipate some of the tools that will appear in later sections.

\vspace{.5em}

\noindent a) Identify the generators of $SL(2,\mathbb{C})$ that stabilizes the null ray in momentum space
\be
 p^\mu=\alpha q^\mu(0,0),~~~\alpha>0.
\ee
Write down the commutation relations for this subgroup.
\vspace{.5em}

\noindent b) Starting from the helicity basis states~\eqref{hb} let us construct the superposition
\be
|\Delta,0,0;\ell\rangle=\int_0^\infty d\omega \omega^{\Delta-1}|\omega,0,0;\ell\rangle.
\ee
How do the generators of the subgroup in part a) act on this state?  Evaluate the inner product $\langle \Delta',z',\bz';\ell'|\Delta,z,\bz;\ell\rangle$ for $\Delta,\Delta'\in 1+i\mathbb{R}$.
\vspace{.5em}

\noindent c) Now consider the action of the remaining Poincar\'e generators on the state $|\Delta,0,0;\ell\rangle$. We call a state quasi-primary if it is an eigenstate of $L_0$ and $\bar{L}_0$ that is annihilated by $L_1$ and $\bar{L}_1$.  For what values of $\Delta$ are there other states in this multiplet that are also quasi-primary?
\vspace{.5em}

\noindent d) Write down the corresponding position space wavefunction $\phi_{\Delta}(X;w,\bw)$ for the $\ell=0$ case.  How does it behave at large $r$ and fixed $u$.  Repeat for fixed $v$.  Show that under
\be
X^\mu\mapsto \Lambda^\mu_{~\nu}X^\nu,~~~w\mapsto \frac{aw+b}{cw+d}
\ee
the wavefunction transforms as
\be
\phi_{\Delta}\Big(\Lambda^{\mu}_{~\nu} X^\nu;\frac{a w+b}{cw+d},\frac{{\bar a} \bw+{\bar b}}{{\bar c}\bw+{\bar d}}\Big)=|cw+d|^{2\Delta}\phi_{\Delta}(X^\mu;w,\bw).
\ee
Can you write a wavefunction with the same transformation law that satisfies the massive Klein-Gordon equation?

\vspace{.5em}

\section{Symmetry Enhancements}\label{asymptotic}
We are now ready to see how the symmetry algebras we studied in the previous section get enhanced when we include gravity. The book~\cite{Strominger:2017zoo} is a comprehensive reference for the IR investigations leading to celestial holography. Here we will focus on the gravitational examples and follow the conventions in~\cite{Pasterski:2019ceq}.

\subsection{Asymptotic Symmetries}\label{ASG}

Let us start by writing the Poincar\'e generators in the Bondi coordinates $(u,r,z,\bz)$ introduced in~\eqref{ubondi}.  We have
\be\badat{3}\label{xiyf}
\xi&=(1+\frac{u}{2r})Y^z\p_z-\frac{u}{2r}D^\bz D_z Y^z\p_\bz-\frac{1}{2}(u+r)D_zY^z\p_r +\frac{u}{2}D_zY^z\p_u+c.c.\\
&~~~~+f\p_u-\frac{1}{r}(D^zf \p_z+D^\bz f \p_\bz)+D^zD_z f\p_r
\eadat
\ee
where
\be
f_0=1,~~f_1=\frac{z+\bz}{1+z\bz},~~f_2=\frac{i(\bz-z)}{1+z\bz},~~f_3=\frac{1-z\bz}{1+z\bz}
\ee
are the global translations,
\be\begin{array}{lll}\label{lorentz}
   Y^z_{12}=iz,~~  & Y^z_{13}=-\frac{1}{2}(1+z^2),~~ & Y^z_{23}=-\frac{i}{2}(1-z^2) \\
   Y^z_{03}=z,~~  & Y^z_{02}=-\frac{i}{2}(1+z^2),~~ & Y^z_{01}=-\frac{1}{2}(1-z^2) \\
\end{array}
\ee
are the global Lorentz transformations, and $D_A$ are covariant derivatives on the unit $S^2$.  Now we recall from our discussion of conformal isometries of the 2-plane, that in that context there was a natural generalization from global $SL(2,\mathbb{C})$ generators to local conformal killing vectors.  To check that the complex $w$ plane we embedded in the lightcone via~\eqref{qref} naturally maps to the celestial sphere, we note that the future light cone of the origin in Bondi coordinates is given by $u=0$.  The large-$r$ limit can be thought of as a different choice of cross section.\footnote{Since null infinity is also the lightcone of spatial infinity, we can also view the canonical celestial sphere as the intersection of this lightcone of the origin and the lightcone of spatial infinity.} It is quick to check that the $\xi(Y)$ is tangent to this lightcone.  Meanwhile we note that the $r\p_r$ term is what gives us an isometry in 4D as opposed to only a conformal isometry in 2D for the case of boosts, where $D_AY^A\neq0$.

So what would happen if we tried to generalize $Y^z=Y^z_{ij}$ to $Y^z(z)$? Starting from the flat metric and taking a Lie derivative we find
\be\label{shift}
\mathcal{L}_{\xi_Y}\eta=-ru(D_z^3 Y^zdz^2+D_\bz^3 Y^\bz d\bz^2)
\ee
where we have dropped possible contact terms that appear in the round sphere metric when $Y$ has poles.  We see that this vanishes at $u=0$, so up to the aforementioned punctures, we have an isometry of the lightcone, but not an isometry of $\mathbb{R}^{1,3}$.

However, we will find that there is still a place for these symmetries when we include dynamical gravity in 4D.  The previous section explored the equivalence between isometries of $\mathbb{R}^{1,d+1}$ and conformal isometries of $\mathbb{R}^{d}$ focusing on the $d=2$ case.  By only demanding that we stayed in the same conformal class of metrics, we found more symmetries in $d$-dimensions than we would have had if we had looked at isometries of $\mathbb{R}^{d}$ instead.  We will now see that in the Lorentzian case there is a natural class of metrics corresponding to asymptotically flat spacetimes.  The requirement that our diffeomorphism preserve this class relaxes~\eqref{lieder} and gives us a much larger class of `asymptotic symmetries.'

The set-up is as follows.  We want to study the phase space of solutions to Einstein's equations with vanishing cosmological constant and localized stress tensor sources
\be\label{eom}
R_{\mu\nu}-\frac{1}{2}g_{\mu\nu}R=8\pi G T^M_{\mu\nu}.
\ee
To study gravitational radiation, we will want to push our incoming and outgoing Cauchy surfaces to past and future null infinity.  Now to have a well defined initial value problem we need to fix our gauge. 
 The process of identifying the asymptotic symmetry group is as follows:
\begin{itemize}
    \item pick a gauge that will be convenient for analyzing the behavior of the metric near the null boundaries $\mathcal{I}^\pm$;
    \item identify the physically relevant falloffs and the free data that specifies a given solution and will coordinatize our phase space;
    \item identify any residual diffeomorphisms that act non-trivially on this data. 
\end{itemize}
The asymptotic symmetry group is
\be\label{asg}
\mathrm{Asymptotic~Symmetries}=\frac{\mathrm{Allowed~Symmetries}}{\mathrm{Trivial~Symmetries}}
\ee
namely those allowed residual diffeomorphisms that act non-trivially at the boundary data.

This procedure was carried out by Bondi, van der Burg, Metzner, and Sachs~\cite{Bondi:1962px,Sachs:1962wk,Sachs:1962zza}. After gauge fixing the metric near future null infinity takes the form
\be\badat{3}\label{bondi}
ds^2=&-du^2-2dudr+2r^2\gamma_{z\bz}dzd\bz\\
&+\frac{2m_B}{r}du^2+rC_{zz}dz^2+rC_{\bz\bz}d\bz^2\\
&+\left[(D^zC_{zz}-\frac{1}{4r}D_z(C_{zz}C^{zz})+\frac{4}{3r}N_z)dudz+c.c.\right]+...
\eadat\ee
while the physical matter falloffs are 
\be\badat{3}\label{matter}
T^M_{uu}\sim\mathcal{O}(r^{-2}),~~T^M_{ur}\sim\mathcal{O}(r^{-4}),~~T^M_{rr}\sim\mathcal{O}(r^{-4})\\
T^M_{uA}\sim\mathcal{O}(r^{-2}),~~T^M_{rA}\sim\mathcal{O}(r^{-3}),~~T^M_{AB}\sim\mathcal{O}(r^{-1}).\\
\eadat\ee
We see that the leading behavior matches the flat metric in retarded radial coordinates~\eqref{ubondi2}.
Solutions are specified by 
\be\label{free}
\{m_B,N_z,C_{zz}\}
\ee
where $m_B$ is the Bondi mass, $N_z$ is the angular momentum aspect and $C_{zz}$ captures the radiative data with $N_{zz}=\p_u C_{zz}$ referred to as the news tensor.  We further note $C_{\bz\bz}=(C_{zz})^*$ and $N_{\bz}=(N_{z})^*$.
One can already see from the orders written explicitly in~\eqref{bondi} that the subleading orders of the metric are solved for in terms of the leading data.
This is done recursively in the large-$r$ expansion using~\eqref{eom} and gauge fixing conditions. You will have a chance to go through this procedure for the simpler electromagnetic case in the exercises. In Bondi gauge, this radial coordinate parameterizes outgoing null geodesics, as can be seen by setting $du=dz=d\bz=0$ in~\eqref{bondi}.

While each of the quantities in~\eqref{free} vary as a function a function of $(u,z,\bz)$, only $C_{zz}$ can be freely specified as a function along $\mathcal{I}^+$.  The $u$ evolution of $m_B$ and $N_z$ are constrained.  This is as expected in a system with gauge invariance.  Our metric falloffs are such that the conformal boundary has the same causal structure as in the Minkowski case considered in the last section and illustrated in figure~\ref{penrose}.  While $u$ is a timelike coordinate for fixed finite $r$, as we limit to the conformal boundary, we are taking a null Cauchy surface whose normal vector in the flat limit is
\be n^\mu\p_\mu=\p_u-\frac{1}{2}\p_r.\ee
Because of the falloffs~\eqref{bondi} and~\eqref{matter}, the constraint equations that specify $n^\mu G_{\mu\nu}$  reduce to
\be\label{constraints}
\p_u m_B=\frac{1}{4}[D_z^2N^{zz}+D_\bz^2 N^{\bz\bz}]-T_{uu},~~~\p_u N_z=\frac{1}{4}D_z[D_z^2C^{zz}-D_\bz^2 C^{\bz\bz}]+D_zm_B-T_{uz}
\ee
where we have grouped terms quadratic in the metric into the $T_{\mu\nu}$ with the matter contributions
\be\badat{3}
T_{uu}&\equiv \frac{1}{4}N_{zz}N^{zz}+4\pi G\lim\limits_{r\rightarrow\infty}r^2T^M_{uu}\\
T_{uz}&\equiv -\frac{1}{4}D_z[C_{zz}N^{zz}]-\frac{1}{2}C_{zz}D_z N^{zz}+8\pi G\lim\limits_{r\rightarrow\infty}r^2T^M_{uz}.
\eadat\ee
We see that we have specified our solution once we give $C_{zz}(u,z,\bz)$ and the initial values of $m_B$ and $N_A$ in the $u\rightarrow-\infty$ limit. These initial values will appear in the canonical charges for our asymptotic symmetries.

With our class of asymptotically flat metrics in hand, as well as a better understanding of the corresponding phase space, let us now return to the question of asymptotic symmetries. The Minkowski metric is quite special. For a generic metric within the class of asymptotically flat spacetimes, we would not expect any isometries.  The desire to recover some notion of Poincar\'e transformations acting on spacetimes which `look flat' at large distance scales, forces us to generalize to some notion of asymptotic symmetries.
The perhaps surprising outcome is that one lands on a much larger symmetry group. 

The $r^{-n}$ falloffs of the metric~\eqref{bondi} are preserved by residual diffeomorphisms whose leading behavior matches~\eqref{xiyf} but where now we promote the  $f_i$ and $Y_{ij}^z$ to free functions
\be
\{f(z,\bz),~~Y^z(z)\}.
\ee
One can verify this by taking a lie derivative of~\eqref{bondi} along~\eqref{xiyf} with these parameters. The former are called {\it supertranslations}, the latter {\it superrotations}. The vector fields~\eqref{xiyf} will get subleading corrections that depend on the gauge fixing used.  The inhomogeneous shifts~\eqref{shift} and similar ones for supertranslations
\be
\delta g_{uz}=-D_z(1+D^zD_z )f,~~~\delta g_{zz}=-2rD_z^2 f
\ee
are no longer a problem, since they respect the large-$r$ falloffs.\footnote{Some care is needed to enlarge the standard phase space to allow the early and late $u$ behavior associated to superrotations as a gauge symmetry~\cite{Barnich:2011mi,Kapec:2014opa,Strominger:2016wns}.}  The superrotations were excluded from earlier asymptotic symmetry analyses since, beyond the global Lorentz transformations~\eqref{lorentz}, they will be singular at isolated points on the celestial sphere.  This is not a deterrent for us since, after all, we want to connect this story to a 2D CFT and in that context the local symmetry algebras are known to play an important role.

The next step is to compute the canonical charges for these asymptotic symmetries.  A good pedagogical reference on this subject is~\cite{Compere:2018aar}.
Although the Einstein-Hilbert action is small-diffeomorphism invariant, there will be non-trivial boundary terms for large diffeomorphisms. The corresponding charges are codimension-2, defined in terms of the data at the boundary of a given Cauchy surface.
For the purpose of these lectures it will suffice to quote the results for the linearized charge that we will need in what follows
\be\label{Qfy}
Q^+[f,Y]=\frac{1}{8\pi G}\int_{\mathcal{I}^+_-}\left[2m_B(f+\frac{u}{2}D_AY^A)+Y^A N_A\right].
\ee
The $+$ superscript is to indicate this charge is defined at future null infinity. Here we use $\mathcal{I}^+_-$ to denote the past limit $u\mapsto-\infty$ of future null infinity, which we regard as the spatial boundary of our Cauchy surface that hugs future null infinity.\footnote{This charge generates the corresponding asymptotic symmetry transformation for the data on our Cauchy surface. However, unlike more familiar conserved charges, if we take a slice through our spacetime that is not a Cauchy slice but instead asymptotes to a different $u$-cut, the value of $Q_{cut}$ will change in a manner determined by the constraint equations~\eqref{constraints}.}

At this stage we have focused on future null infinity.  To have a well defined scattering problem we need to understand how to match the action of these symmetries on future null infinity to analogous ones at past null infinity.  We will come to this point in section~\ref{ward}. To demonstrate that the $\mathcal{S}$-matrix indeed obeys a Ward identity for these symmetries we will need to review another facet of long-distance/low-energy scattering dynamics that was developed contemporaneously to the asymptotic symmetry story of BMS: soft theorems.

\subsection{Soft Theorems}\label{softthm}

Around the same time BMS were studying asymptotic symmetries, Weinberg and others were examining the low energy limit of scattering~\cite{Weinberg:1965nx,Low:1954kd,Gell-Mann:1954wra,Low:1958sn,Burnett:1967km}.  Here we will focus on the gravitational case, reviewing the derivation for the leading Weinberg soft theorem and giving its extension to subleading order~\cite{Cachazo:2014fwa}. The leading photon case is covered in depth in the textbook~\cite{Weinberg:1995mt}.

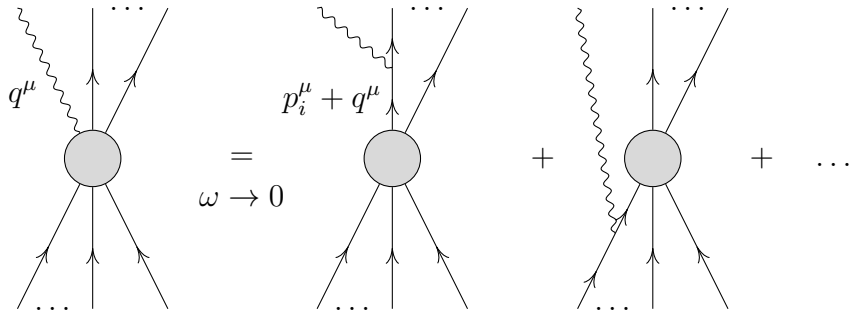
\begin{figure}[b!]
\centering
\begin{tikzpicture}
   \begin{feynman}[every blob={/tikz/fill=gray!30,/tikz/inner sep=2pt}]
\vertex at (-1,-2)  (a) ;
\vertex at (0,-2)  (b) ;
\vertex at (1,-2)  (c) ;
\vertex at (-1,2)  (d) ;
\vertex at (0,2)  (e) ;
\vertex at (1,2)  (f) ;
 \node at (-.5,-2) {\ldots};
  \node at (.5,2) {\ldots};
  \node at (1,0) {};
\vertex[blob,label={}] (m) at ( 0, 0) {};
\diagram*{
(a) -- [fermion2] (m) -- [boson, edge label={$q^\mu$}] (d);
(b) -- [fermion2] (m) -- [fermion2] (e);
(c) -- [fermion2] (m) -- [fermion2] (f);
};
  \end{feynman}
\end{tikzpicture}
\begin{tikzpicture}
   \begin{feynman}[every blob={/tikz/fill=gray!30,/tikz/inner sep=2pt}]
\vertex at (-1,-2)  (a) ;
\vertex at (0,-2)  (b) ;
\vertex at (1,-2)  (c) ;
\vertex at (-1,2)  (d) ;
\vertex at (0,2)  (e) ;
\vertex at (1,2)  (f) ;
\vertex at (0,1.2)  (g) ;
 \node at (-.5,-2) {\ldots};
  \node at (.5,2) {\ldots};
    \node at (-2,0) {$=$};
       \node at (-2,-.5) {$\omega\rightarrow 0$};
        \node at (2,0) {$+$};
\vertex[blob,label={}] (m) at ( 0, 0) {};
\diagram*{
(a) -- [fermion2] (m);
(b) -- [fermion2] (m) -- [fermion2,edge label={$p_i^\mu+q^\mu$}] (g) -- [fermion2] (e);
(c) -- [fermion2] (m) -- [fermion2] (f);
(g) -- [boson] (d);
};
  \end{feynman}
\end{tikzpicture}
\begin{tikzpicture}
   \begin{feynman}[every blob={/tikz/fill=gray!30,/tikz/inner sep=2pt}]
\vertex at (-1,-2)  (a) ;
\vertex at (0,-2)  (b) ;
\vertex at (1,-2)  (c) ;
\vertex at (-1,2)  (d) ;
\vertex at (0,2)  (e) ;
\vertex at (1,2)  (f) ;
\vertex at (-.5,-1)  (g) ;
 \node at (-.5,-2) {\ldots};
  \node at (.5,2) {\ldots};
        \node at (2,0) {$+~~~~$\dots};
\vertex[blob,label={}] (m) at ( 0, 0) {};
\diagram*{
(a) -- [fermion2] (g) -- [fermion2] (m);
(b) -- [fermion2] (m) -- [fermion2] (e);
(c) -- [fermion2] (m) -- [fermion2] (f);
(g) -- [boson] (d);
};
  \end{feynman}
\end{tikzpicture}
\caption{In the limit where the energy of a graviton or gauge boson goes `soft,' the leading contribution will come from attaching to each of the `hard' external legs.
\label{soft theorem}}
\end{figure}

Let us start with gravity coupled to scalar matter~\cite{Weinberg:1965nx}.   Consider the case illustrated in figure~\ref{soft theorem}, where an outgoing graviton's energy is taken soft.  When this graviton attaches to the $i^{th}$ external leg, we get an extra propagator that goes like
\be
\frac{-i}{(p_i+\eta_i q)^2+m_i^2-i\epsilon}=\frac{-i}{2p_i\cdot q-i\epsilon}
\ee
where $\eta_i=\pm1$ depending on whether the particle is outgoing or incoming.  Here we have used the fact that $p_i$ is on-shell. The difference between massless and massive scattering is that for the latter the denominator can vanish in the collinear limit, otherwise we can drop the $i\epsilon$.  This quantity scales like $\omega^{-1}$.  Adding the vertex factors we get
\be
\frac{\kappa}{4}\frac{(2p_i^\mu+\eta_i q^\mu)(2p_i^\nu+\eta_i q^\nu)}{2p_i\cdot q-i\epsilon}\rightarrow \frac{\kappa}{2}\frac{\eta_i p_i^\mu p_i^\nu}{p_i\cdot q}
\ee
as we send $\omega\rightarrow0$.  This factor multiplies the rest of the Feynman diagram, which matches that of the scattering process with this graviton removed. Other Feynman diagrams that can contribute graviton emission will be subleading to this $\mathcal{O}(\omega^{-1})$ pole.

While the Feynman rules for external legs of different spins will change the form of the propagator, vertex, and external state contraction, the final form of the soft factorization theorem takes a universal form.  In~\cite{Cachazo:2014fwa} it was shown how to extend this to subleading order in $\omega$.  Together we have
\be\label{sfthm}
\langle out|a_{\pm}(q)\mathcal{S}|in\rangle=(S^{(0)\pm}+S^{(1)\pm})\langle out|\mathcal{S}|in\rangle+\mathcal{O}(\omega)
\ee
where
\be
S^{(0)\pm}=\frac{\kappa}{2}\sum_i\eta_i\frac{(p_i\cdot \epsilon^\pm)^2}{p_k\cdot q},~~~S^{(1)\pm}=-i\frac{\kappa}{2}\sum_i\eta_i\frac{p_{i\mu}\epsilon^{\pm\mu\nu}q^\lambda J_{i\lambda\nu}}{p_i\cdot q}.
\ee
While the leading soft theorem is a multiplicative constant in momentum space, the subleading soft theorem involves the angular momentum operator which will act as a differential operator in this basis.  For instance spinor helicity variable this takes the form
\be
J_{\mu\nu}\sigma^{\mu}_{a{\dot a}}\sigma^{\nu}_{b{\dot b}}=-2J_{ab}\varepsilon_{\dot a\dot b}-2\varepsilon_{ab}\bar{J}_{\dot a\dot b}
\ee
where
\be
J_{ab}=\frac{i}{2}\left(\lambda_a\frac{\p}{\p \lambda^b}+\lambda_b\frac{\p}{\p \lambda^a}\right),~~~\bar{J}_{\dot a\dot b}=\frac{i}{2}\left(\bar{\lambda}_{\dot a}\frac{\p}{\p \bar{\lambda}^{\dot b}}+\bar{\lambda}_{\dot b}\frac{\p}{\p \bar{\lambda}^{\dot a}}\right)
\ee
and similarly for the anti-chiral part. We will return to this point in the following section when we map this soft operator to a putative 2D stress tensor.

Before moving on towards the asymptotic symmetry Ward identities we are after, we should point out various insights this soft limit provides.  First, using what we noted in section~\ref{momentum} about gauge transformations and the polarization tensors shifting along $\epsilon_\pm^\mu\mapsto\epsilon_\pm^\mu+\alpha q^\mu$, we see that if we take
\be
\epsilon_\pm^\mu\epsilon_\pm^\nu\mapsto \epsilon_\pm^\mu q^\nu+ q^\mu \epsilon_\pm^\nu 
\ee
we have
\be
S^{(0),\pm}\mapsto \frac{\kappa}{2}\sum_i\eta_i p_i\cdot\epsilon^\pm,~~~S^{(1),\pm}\mapsto-i \frac{\kappa}{2}\sum_i\eta_i \epsilon^{\pm \nu}q^\lambda J_{i\lambda\nu}
\ee
which vanish due to total momentum and angular momentum conservation, respectively.  This is also a manifestation of the equivalence principle.  

Furthermore, there is no reason to stop at single emissions. Indeed soft radiation can be seen to be unavoidable.  So long as the external legs are charged under a given gauge field, such as gravity, there will be virtual exchanges.  These loop contributions are IR divergent in 4D and exponentiate in a manner that makes the exclusive amplitude vanish. The standard protocol is to introduce a detector cutoff so that inclusive quantities which allow unmeasured virtual soft emissions are finite.  From the perspective of the soft physics program these IR divergence issues can be seen as symptoms of asymptotic charge non-conservation~\cite{Kapec:2017tkm}.  There is also a natural reason make measurements in the deep infrared in this context.  Such observables are called memory effects and they require specially adapted non-calorimetric detector setups~\cite{Strominger:2014pwa,1502.06120PSZ,Pasterski:2015zua,Susskind:2015hpa}.

\subsection{Ward Identities}\label{ward}

We now have almost all the elements we need to demonstrate that the perturbative gravitational $\mathcal{S}$-matrix obeys a Ward identity for the asymptotic symmetries we studied in section~\ref{ASG} and, moreover, that these Ward identities are equivalent to the soft theorems of section~\ref{softthm}. The final ingredient is an insight by Strominger~\cite{Strominger:2013jfa} that we can impose an antipodal matching condition across spatial infinity.

If we repeat the computations in section~\ref{ASG} at past null infinity, we will have charges defined in terms of data at $\mathcal{I}^-_+$ the $v\rightarrow+\infty$ limit of $\mathcal{I}^-$.  On the Penrose diagram of figure~\ref{penrose} the past limit of future null infinity and the future limit of past null infinity both look like they are at the same point $i^0$.  However, these are actually separated by an infinite amount of time. Motivated by CPT invariance, Strominger proposed the matching
\be\label{antipodal}
    C_{zz}|_{\mathcal{I}^+_-}=    C_{zz}|_{\mathcal{I}^-_+},~~~m_B|_{\mathcal{I}^+_-}=    m_B|_{\mathcal{I}^-_+},~~~\p_{[z}N_{\bz]}|_{\mathcal{I}^+_-}=  \p_{[z}N_{\bz]}|_{\mathcal{I}^-_+}
\ee
in coordinates, as in section~\ref{nullinfty}, where the $z$ coordinates near $\mathcal{I}^\pm$ incorporate this antipodal matching.  This is consistent with how the Poincar\'e generators act, as well as the soft theorem-based derivation of the Ward identities we will give below. You will get a chance to explore this in the simple case of the electromagnetic field of a moving point charge in the exercises. A systematic treatment for the case of Maxwell theory can be found in~\cite{Prabhu:2018gzs}.

Comparing~\eqref{antipodal} to~\eqref{Qfy}, we see that with an appropriate identification of the $f(z,\bz)$ and $Y^z(z)$ corresponding to the diagonal subgroup of $BMS^+\times BMS^-$, the classical charge conservation between the incoming and out-going states is $Q^+[\xi]=Q^-[\xi]$.  Because these charges generate transformations on the incoming and outgoing states separately, the corresponding Ward identity we would like to hold for the $\mathcal{S}$-matrix is 
\be\label{wardid}
\langle out | Q^+[\xi] \mathcal{S}-\mathcal{S}Q^-[\xi]|in\rangle.
\ee
We can show that this indeed is observed in perturbative gravitational scattering using the soft theorem~\eqref{sfthm} and the constraint equation~\eqref{constraints}.  We will focus on the subleading soft graviton case here. Using the simple observation that 
\be
\int_{\mathcal{I}^+} du \p_u \left(\cdot\right)= \left.\left(\cdot\right)\right|^{\mathcal{I}^+_+}_{\mathcal{I}^+_-}
\ee
combined with the fact that the $u$-derivatives of the Bondi mass and angular momentum aspect are given by the constraint equations, we can write the charges defined near $i^0$ in terms of an integral along $u$ and a contribution near $i^+$.  This future timelike infinity contribution will capture the kinematics of the massive external states.  The case of only massless scattering simplifies things a bit and we are left with 
\be
Q^+[Y]=-\frac{1}{8\pi G}\int_{\mathcal{I}^+}\sqrt{\gamma}d^2z du \left[\frac{1}{2}D_z^3 Y^z u\p_u C^{zz}+Y^z T_{uz}+D_zY^z uT_{uu}\right]
\ee
where we have complexified our transformation so that $Y^\bz=0$, which lets us isolate the relations coming from each helicity soft theorem.
The first term can be evaluated using the saddle point approximation~\eqref{eq:hatH} and then the subleading soft theorem, by noting that the $u$-integral picks out $\mathcal{O}(\omega^0)$ part of the soft limit.
Meanwhile on energy eigenstates the stress tensor terms, which we will call the hard charge $Q_H^+[Y]$, act as
\be\label{Qh}
\langle\omega_k,z_k,\bz_k|Q_H^+[Y]=i\left( Y(z_k)\p_{z_k}-\frac{1}{2}D_{z_{k}}Y^{z_{k}}\omega_k\p_{\omega_k}\right)\langle\omega_k,z_k,\bz_k|.
\ee
The soft theorem gets contributions from both incoming and outgoing external legs, while the hard charges act on the $in$ and $out$ states separately.  Together with their incoming analogs, we find~\eqref{wardid} holds~\cite{Kapec:2014opa}.

This splitting of the asymptotic symmetry generators persists to the more general case
\be
Q^\pm[\xi]=Q_S^\pm[\xi]+Q_H^\pm[\xi]
\ee
where the soft charge is linear in the metric/gauge field and generates the inhomogeneous shift as in~\ref{shift}.  Namely, this mode is symplectically paired to the Goldstone mode of the corresponding asymptotic symmetry. This soft operator is thus a harbinger of the spontaneous symmetry breaking of the asymptotic symmetry generators that are not part of the subgroup isomorphic to Poincar\'e that leaves invariant a given vacuum.  In $\mathcal{S}$-matrix elements this is realized as an extra soft gauge boson insertion. Diagonalizing this operator requires going to coherent dressed states. Its classical value corresponds to so-called memory effect observables.  In celestial CFT, this operator will give rise to currents.

Let's see how this last feature presages our efforts in the next section.  We see from~\eqref{Qh} that to diagonalize the action of the hard charge on the matter particles, we will need to go to Rindler energy eigenstates. We then have a candidate stress tensor~\cite{Kapec:2016jld}
\be\label{tzz}
T_{zz}=-i\frac{3!}{8\pi G}\int  d^2 w \frac{1}{(z-w)^4}\int du u\p_u C^w_{~\bw},
\ee
constructed from this soft mode when $Y^w=\frac{2i}{z-w}$. The soft theorem becomes
\be\label{Tward}
\langle T_{zz}\mathcal{O}_1...\mathcal{O}_n\rangle=\sum_k\left[\frac{h_{k}}{(z-z_k)^2}+\frac{\p_{z_k}}{z-z_k}\right]\langle\mathcal{O}_1...\mathcal{O}_n\rangle
\ee
and similarly for $\bar{T}_{\bz\bz}$ where
\be
h_k=\frac{1}{2}\left(\ell_k-\omega_k\p_{\omega_k}\right),~~~\bar{h}_k=\frac{1}{2}\left(-\ell_k-\omega_k\p_{\omega_k}\right)
\ee
and we have suggestively used the map~\eqref{map}.
We see that in the boost basis the Ward identity for superrotations in 4D looks like the Ward identity for a stress tensor in 2D. 

While we have focused on the gravitational case in this section, the gauge theory analog of the leading soft theorem gives a celestial Kac-Moody symmetry~\cite{He:2015zea}.  This points to a 2D description being the most natural way to arrange our infinite dimensional symmetry enhancements. Before proceeding to follow this route and introduce celestial amplitudes, let us close this section with a couple comments on what we have already gained from this soft physics story.  We mentioned in the previous subsection that the asymptotic symmetry perspective gives us another way to interpret IR divergences appearing in amplitudes.  We would also like to be able to take advantage of the infinite number of extra symmetries we have at hand. For example, these symmetries should constrain black hole evaporation, where the horizon analog amounts to a source of soft hair~\cite{Hawking:2015qqa,Hawking:2016msc,Hawking:2016sgy,Strominger:2017aeh}.  While our proof of the Ward identity has used perturbation theory, the real power comes from asserting that these symmetries will exist in the full quantum theory of gravity.  The main aim of going to the celestial basis is to put ourselves in the best position to take advantage of these symmetries.

\vspace{1em}
 
\noindent{\bf Exercise:} The leading soft photon theorem is isomorphic to a Ward identity for U(1) gauge transformations that are order $\mathcal{O}(r^0)$ at null infinity. \vspace{.5em}

\noindent a) Starting from the radial expansion of the vector potential $A_\mu$ 
\be
A_\mu=\sum_n  r^{-n} A^{(n)}_{\mu}+r^{-n}\log r A^{(n),\log}_{\mu}
\ee
write out the radial expansion for Maxwell's equations and the harmonic gauge condition. What is the free data?\vspace{.5em}

\noindent b) Consider the solution sourced by a constantly moving particle with charge Q and four velocity $U^\mu=\gamma(1,\vec{\beta})$
\be
j_\mu=Q\int d\tau U_{\mu}\delta^{(4)}(x^\nu-U^\nu\tau)
\ee
where  $\gamma^2=(1-\beta^2)^{-1}$.
Write down the corresponding Li\'enard-Wiechert potential $A_\mu$.  How does it behave at large $r$ fixed $u$, respectively fixed $v$?  How does it behave if you then take $u\rightarrow-\infty$, respectively $v\rightarrow+\infty$?
\vspace{.5em}

\noindent c) Write down the canonical charge for this gauge transformation as an integral over spatial infinity.  Then use Stokes' theorem and the constraint equation to write this as an integral of the gauge field and matter current over null infinity. \vspace{.5em}

\noindent d) Show that the leading soft photon theorem is equivalent to the semi-classical Ward identity for these large gauge transformations 
\be
\langle out | Q^+[\lambda] \mathcal{S}-\mathcal{S}Q^-[\lambda]|in\rangle.
\ee
Discuss the role of the antipodal matching condition.

\section{The Holographic Map}\label{dictionary}

We've see that infinite dimensional symmetry enhancements are encoded in the infrared behavior of amplitudes and seem to have a natural description in terms of currents in a 2D celestial CFT.  We've also seen from the stress tensor example a hint of what scattering basis we want to go to.  We will now start by constructing the dictionary for single particle external states, and then explore how this holographic map rearranges features of scattering familiar from momentum space into a CFT language.

\subsection{The Start of a Dictionary}\label{dict}

While the IR dynamics hinted at the presence of 2D currents, the aim of~\cite{Pasterski:2016qvg,Pasterski:2017kqt,Pasterski:2017ylz} was to take us beyond the soft limit of scattering.  The way to do this is to change our scattering basis.  This amounts to changing the wavepackets we prepare for the in and out states. Let us start with the definition of a conformal primary wavefunction~\cite{Pasterski:2017kqt}.
\vspace{1em}
 
\noindent{\bf Definition:} A {\it conformal primary wavefunction} is a function on $\mathbb{R}^{1,3}$ which transforms under $SL(2,\mathbb{C})$ as a 2D conformal primary of conformal dimension $\Delta$ and spin $J$, and a 4D (spinor-) tensor field of spin-$s$:
\begin{equation}\label{Defgenprim}
    \badat{2}
\Phi^{s}_{\Delta,J}\Big(\Lambda^{\mu}_{~\nu} X^\nu;\frac{a w+b}{cw+d},\frac{{\bar a} \bw+{\bar b}}{{\bar c}\bw+{\bar d}}\Big)=(cw+d)^{\Delta+J}({\bar c}\bw+{\bar d})^{\Delta-J}D_s(\Lambda)\Phi^{s}_{\Delta,J}(X^\mu;w,\bw)\,,
\eadat
\end{equation}
where $D_s(\Lambda)$ is the 3+1D spin-$s$ representation of the Lorentz algebra. \vspace{1em}

\noindent In what follows we will also demand that our wavefunctions are solutions to the corresponding spin-$s$ linearized bulk equations of motion.  For massless fields, radiative solutions will have $|J|=s$.  These restrictions can be relaxed in which case we refer to them as {\it generalized primaries}~\cite{Pasterski:2020pdk}. Given such a wavefunction, we can construct a celestial primary operator as follows~\cite{Donnay:2020guq}
 \be\label{qdelta}
\O^{s,\pm}_{\Delta,J}(w,\bw)\equiv i(\hat{O}^{s}(X^\mu),\Phi^s_{\Delta^*,-J}(X_\mp^\mu;w,\bw))_{\Sigma}\,.
\ee
The $\pm$ superscript on the operator indicates whether the operator is incoming or outgoing.  It is selected by an appropriate analytic continuation of $X^\mu$ in the wavefunction,\footnote{We will see in what follows that this regulator is needed for the convergence of the
Mellin transform that defines these wavefunctions, and will also regulate singularities they would otherwise have. We will suppress this $\pm$ when reasonable, as well as the $s$ superscript on operators and wavefunctions whenever our modes are radiative and its value is apparent from the context.}  as indicated by the subscript $X_\pm^\mu=X^\mu\pm i\{-1,0,0,0\}$. Much like in the momentum case, we need a local operator in the bulk and an inner product on a Cauchy slice.  To prepare the in and out states we can push these Cauchy slices to $\mathcal{I}^\pm$ so that we can define these operators in terms of boundary data.  

By construction, the transformation law~\eqref{Defgenprim} gives us an object which transforms like a local 2D quasi-primary.  We can see this as follows.  In section~\ref{isom} we introduced $SL(2,\mathbb{C})$ generators acting as isometries in 4D~\eqref{poincare} and conformal isometries in 2D~\eqref{2con}, explicitly realizing this isomorphism by embedding the celestial sphere into the $\mathbb{R}^{1,3}$ lightcone via~\eqref{qref}. If we distinguish the action on $X^\mu$ versus on $(w,\bw)$ with appropriate subscripts, the infinitesimal version of~\eqref{qdelta} amounts to~\cite{Pasterski:2021fjn}
\begin{equation}
\label{covariance_generators}
(M_{\mathbb{R}^{1,3}}^{\mu\nu}+M_{\mathbb{C}}^{\mu\nu})\Phi^s_{\Delta,J}(X^\sigma;w,\bw)=0. \, 
\end{equation}
Both the left and right hand sides of~\eqref{qdelta} are operators defined on the 4D Hilbert space $\mathcal{H}$.  Acting with the corresponding generators of section~\ref{momentum} which, for the sake of comparing to the notation here, we denote $M_{\mathcal{H}}^{\mu\nu}$, the Lorentz invariance of the inner product and~\eqref{covariance_generators} implies
\be
[M_{\cal H}^{\mu\nu},\hat{O}^s(X)]=-\mathcal{L}_{M_{\mathbb{R}^{1,3}}^{\mu\nu}}\hat{O}^s(X)~~\Rightarrow~~[M_{\cal H}^{\mu\nu},\O^{s,\pm}_{\Delta,J}(w,\bw)]=-\mathcal{L}_{M_{\mathbb{C}}^{\mu\nu}}\O^{s,\pm}_{\Delta,J}(w,\bw)
\ee
up to the same caveat about gauge transformation equivalence classes discussed at the end of section~\ref{momentum}.  Explicitly, we thus have
\be\label{Liaction}
\badat{3}
L_0\mathcal{O}_{h,\bar{h}}=2(w\p_w+h)\mathcal{O}_{h,\bar{h}},~~~&L_{-1}\mathcal{O}_{h,\bar{h}}=\p_w\mathcal{O}_{h,\bar{h}},~~~&L_{+1}\mathcal{O}_{h,\bar{h}}=(w^2\p_w+2wh)\mathcal{O}_{h,\bar{h}}\\
\bar{L}_0\mathcal{O}_{h,\bar{h}}=2(\bw\p_\bw+\bar{h})\mathcal{O}_{h,\bar{h}},~~~&\bar{L}_{-1}\mathcal{O}_{h,\bar{h}}=\p_\bw\mathcal{O}_{h,\bar{h}},~~~&\bar{L}_{+1}\mathcal{O}_{h,\bar{h}}=(\bw^2\p_\bw+2\bw \bar{h})\mathcal{O}_{h,\bar{h}}\\
\eadat
\ee
where
\be
h=\frac{1}{2}(\Delta+J),~~~\bar{h}=\frac{1}{2}(\Delta-J).
\ee

While we have gotten a long way by specifying the covariance properties of the external wavefunctions, let us actually construct them. The building blocks for such conformal primaries are objects which transform with definite $SL(2,\mathbb{C})$ weight under
 \be\label{mobius2}
X^\mu\mapsto \Lambda^\mu_{~\nu}X^\nu\,,~~~ w\mapsto \frac{a w+b}{cw+d}\,,~~~\bw\mapsto \frac{{\bar a} \bw+{\bar b}}{{\bar c}\bw+{\bar d}}\,.
 \ee
 Starting from the reference vector $q^\mu$ and polarization vectors $\epsilon_{\pm}^\mu$ of~\eqref{qref} and~\eqref{polar}, we can construct the following null tetrad ~\cite{Pasterski:2020pdk}
\be\label{tetrad}
l^\mu=\frac{q^\mu}{-q\cdot X}\,, ~~~n^\mu=X^\mu+\frac{X^2}{2}l^\mu\,, ~~~m^\mu=\epsilon^\mu_w+(\epsilon_w\cdot X) l^\mu\,, ~~~\bar{m}^\mu=\epsilon^\mu_\bw +(\epsilon_\bw\cdot X) l^\mu\,,
\ee
which further decomposes into the spin frame

\be\label{spinframe}
l_{a{\dot b}}=o_a\bar{o}_{\dot b}\,,~~n_{a{\dot b}}=\iota_a\bar{\iota}_{\dot b}\,,~~m_{\alpha{\dot b}}=o_a{\bar\iota}_{\dot b}\,,~~\bar{m}_{a{\dot b}}=\iota_a{\bar o}_{\dot b}\,
\ee
where
\be\label{spinframespinors}
o_a=i(-q\cdot X)^{-\frac{1}{2}}|q]_a,~~~\iota_a=\frac{1}{\sqrt{2}}X_{a\dot b}\bar{o}^{\dot b}.
\ee
The $SL(2,\mathbb{C})$ weights of these objects are summarized in table~\ref{table:tetradspinframe}.  Wavefunctions with $\Delta\in\mathbb{C}$ and $J\in\frac{1}{2}\mathbb{Z}$ can be constructed by taking products of these tetrad and spin frame elements with a scalar primary of the form
\be
\varphi^{gen}_{\Delta}\equiv f(X^2)\varphi^\Delta,~~~\varphi^\Delta=\frac{1}{(-q\cdot X)^\Delta}.
\ee

\begin{table}[ht!]
    \centering
    \begin{tabular}{c|c|c|c|c|c|c|c|c}
   & $l^\mu$ & $n^\mu$ & $m^\mu$ & $\bar{m}^\mu$ & $o_a$ & $\bar{o}_{\dot{a}}$ & $\iota_a$ & $\bar{\iota}_{\dot{a}}$   \\ \hline
   $\Delta$ & $0$&$0$ &$0$&$0$&$0$&$0$&$0$&$0$\\
   $J$& $0$&$0$ &$+1$&$-1$&$+\frac{1}{2}$&$-\frac{1}{2}$&$-\frac{1}{2}$&$+\frac{1}{2}$\\
    \end{tabular}
    \caption{$SL(2,\mathbb{C}$) quantum numbers for elements of our tetrad and spin frame.}
    \label{table:tetradspinframe}
\end{table}

 Now to use such wavefunctions in~\eqref{qdelta}, we want them to also satisfy the appropriate linearized equations of motion. We will focus on the integer spin massless case here.  Demanding~\eqref{aheom} we find two sets of solutions
 \be\begin{array}{ll}\label{mellinCPWs}
     A_{\Delta,J=+ 1}=m \varphi^{\Delta} \,,&~~~ A_{\Delta,J=- 1}=\bar{m} \varphi^{\Delta}\,, \\
    h_{\Delta,J=+2}=m m \varphi^{\Delta} \,,&~~~  h_{\Delta,J=-2}=\bar{m} \bar{m} \varphi^{\Delta}\,,
\end{array}
\ee
and
 \be\begin{array}{ll}\label{shadowCPWs}
  \widetilde{A}_{\Delta,J=-1}={-}(-X^2)^{\Delta-1}\bar{m}\varphi^\Delta\,, &~~~\widetilde{A}_{\Delta,J=+1}={-}(-X^2)^{\Delta-1}m\varphi^\Delta\,,\\
     \widetilde{h}_{\Delta,J=-2}\,=(-X^2)^{\Delta-1}\bar{m}\bar{m}\varphi^\Delta\,, &~~~ \widetilde{h}_{\Delta,J=+2}\,=(-X^2)^{\Delta-1}{m} {m}\varphi^\Delta\,.
    \end{array}
\ee
The solutions~\eqref{mellinCPWs} and~\eqref{shadowCPWs} are related by the conformal shadow transform
 \begin{equation}\label{2dShadowTransform}
\widetilde{\O}_{2-\Delta,-J}(w,\bw)=\frac{k_{\Delta,J}}{2\pi} \int d^2w' \frac{\O_{\Delta,J}(w',\bw')}{(w-w')^{2-\Delta-J}(\bw-\bw')^{2-\Delta+J}}\,.
\end{equation}
and are thus not linearly independent.  Here the constant  $k_{\Delta,J}=\Delta-1+|J|$ is chosen so that this transform squares to $(-1)^{2J}$.   We have already actually seen this transform covertly appearing in our definition of the stress tensor~\eqref{tzz}.

A nice feature of~\eqref{mellinCPWs} is that these solutions are gauge equivalent to the Mellin transforms of the corresponding plane wave solutions, up to an overall normalization.   This integral transform is defined as
\begin{equation}\label{Mellin}
\mathcal{M}[f](\Delta)=\int_0^\infty d\omega \omega^{\Delta-1}f(\omega)\equiv \phi(\Delta)\,,
\end{equation}
and can be inverted via
\begin{equation}\label{InverseMellin}
\mathcal{M}^{-1}[\phi](\omega)=\frac{1}{2\pi i}\int_{c-i\infty}^{c+i\infty} d\Delta \, \omega^{-\Delta}\phi(\Delta)=f(\omega)\,.
\end{equation}
For our purposes we can use the contour $c=1$. For the scalar we see that
\be
\frac{1}{(-q\cdot X_\pm)^\Delta}=\frac{(\pm i)^\Delta}{\Gamma(\Delta)}\phi^{\Delta,\pm},~~~\phi^{\Delta,\pm}=\int_0^\infty d\omega \omega^{\Delta-1}e^{\pm i\omega q\cdot X-\omega\epsilon}.
\ee
Meanwhile spin-1 and spin-2 wavefunctions similarly differ by an overall factor and gauge transformation. Explicitly,
\begin{equation}\label{hDelta}\begin{aligned}
A^{\pm}_{\mu;\Delta,J}(X^\mu;w,\bw)
 &=\frac{\Delta-1}{\Delta}\frac{\sqrt 2 (\pm i)^\Delta}{\Gamma(\Delta)}\epsilon_{\mu;J} \phi^{\Delta,\pm}+\nabla_\mu \lambda^{\pm}_{\Delta,J}\,,\\
 h^{\pm}_{\mu\nu;\Delta,J}(X^\mu;w,\bw)
 &=\frac{\Delta-1}{\Delta+1}\frac{(\pm i)^\Delta}{\Gamma(\Delta)} \epsilon_{\mu\nu;J} \phi^{\Delta,\pm}+\nabla_\mu \xi^{\pm}_{\nu;\Delta,J}+\nabla_\nu \xi^{\pm}_{\mu;\Delta,J}\,,
 \end{aligned}
\end{equation}
for gauge parameters $\lambda_{\Delta,J}$, and $\xi^\mu_{\Delta,J}$ whose explicit form can be found in~\cite{Donnay:2020guq}.

By being able to write our wavefunctions as integrals over the on-shell momenta, we can apply these integral transforms directly to $\mathcal{S}$-matrix elements to get our {\it Celestial Amplitudes}.  For the massless case, the map~\eqref{map} becomes
\be\label{mellin}
\langle \mathcal{O}^\pm_{\Delta_1}(z_1,\bz_1)...\mathcal{O}^\pm_{\Delta_n}(z_n,\bz_n)\rangle=\prod_{i=1}^n \int_0^\infty d\omega_i \omega_i^{\Delta_i-1} \langle out|\mathcal{S}|in\rangle.
\ee
This will be the central object of study in what follows. 
Radiative data is captured by $\Delta\in 1+i\mathbb{R}$ on the principal series~\cite{Pasterski:2017kqt}, from which we can analytically continue to $\Delta\in \mathbb{C}$~\cite{Donnay:2020guq}. Similar integral transforms exist for the massive case which are built using tools from the CFT embedding space formalism applied to the $p^2=m^2$ hyperboloid~\cite{Pasterski:2016qvg}.  These will involve a three dimensional integral over the $(y,z,\bz)$ of~\eqref{massive} for each external state, whose kernel strongly weights configurations that are highly boosted towards a chosen reference direction $(w,\bw)$.

\subsection{Soft Limits as Currents}

Now that we have a map that takes us from $\mathcal{S}$-matrix elements in the usual energy eigenstates to boost eigenstates designed to transform as 2D conformal correlators, it is worth revisiting how energetically soft limits are now encoded.  A priori it is not obvious that a factorization that occurs in the limit $\omega\rightarrow 0$ would turn into a similar factorization once we integrate over all energies. 

Let us tackle this problem in a manner that takes the most advantage of the formalism we have reviewed thus far: namely, starting with asymptotic symmetries and our conformal primary wavefunctions.  The operators constructed in~\eqref{qdelta} generate the following shift on the bulk operator $\hat{O}^s(X)$
 \be\label{qshift}
[\O^{s,\pm}_{\Delta,J}(w,\bw),\hat{O}^s(X)]= i\Phi^s_{\Delta,J}(X;w,\bw).
\ee
We thus see that if we can identify conformal primary wavefunctions that correspond our asymptotic symmetries, we know the corresponding celestial basis soft operator/current.  
It turns out that there are only a discrete set of conformal dimensions $\Delta$ for which the conformal primaries are pure gauge.  These are listed for $s\le 2$ in table~\ref{table:Goldstone} below.  Their asymptotic symmetry interpretation can be verified by performing a large-$r$ expansion of the corresponding Goldstone modes.  

\vspace{1em}
\begin{table}[bh!]
\renewcommand*{\arraystretch}{1.3}
\centering
\begin{tabular}{|c|c|c|cc|}
\hline
  & \multicolumn{1}{c|}{${A}^{\Delta}_{\mu}$} & \multicolumn{1}{c|}{$\chi^{\Delta}_{\mu}$}  & \multicolumn{2}{c|}{${h}^{\Delta}_{\mu\nu}$}\\
  \hline
 $\Delta$ &1 &$\frac{1}{2}$ &  1&  0 \\
 symmetry & large $U(1)$ & {large} SUSY  & supertranslation & shadow superrotation $\in$ Diff$(S^2)$\\
\hline
\end{tabular}
\caption{Conformal Goldstone modes of spontaneously broken asymptotic symmetries for particles with spin $1\le s\le2$.  
}
 \label{table:Goldstone}
\end{table}

Let us now see how this bears out in the transformed amplitude~\eqref{mellin}.  As observed in~\cite{Cheung:2016iub}, poles in $\omega$ will translate to poles in $\Delta$.  In particular, if our momentum space amplitude behaves as
\be\label{aexp}
A:=\langle out|\mathcal{S}|in\rangle\sim\omega^{-1} A^{(1)}+A^{(0)}+...
\ee
as we take the energy of one of the scattering states soft, we see that the IR contribution to the Mellin integral 
\be
\int_0^{\Lambda}d\omega\omega^{\Delta-1}\omega^{a}=\frac{\Lambda^{\Delta+a}}{a+\Delta}
\ee
will give poles at $\Delta=1,0,...$ on and to the left of the principal series.   We will return to the UV behavior in section~\ref{moreccft}. So long as it falls off fast enough we can use
\be
\lim\limits_{\epsilon\rightarrow0}\frac{\epsilon}{2}\omega^{\epsilon-1}=\delta(\omega)
\ee
to extract the residue in the limit~\cite{Pate:2019mfs}
\be\label{alim}
\lim\limits_{\Delta\rightarrow-n}(\Delta+n)\int_0^\infty d\omega \omega^{\Delta-1}\sum_k\omega^k A^{(k)}=A^{(n)}.
\ee
We see that we have a conformal basis analog of our result from section~\ref{ward}.  Namely, amplitudes in the conformal basis obey a conformally soft factorization which appear as poles in the celestial amplitude at special values of $\Delta$.  Moreover, we have the nice feature that subleading soft theorems that occur at different orders in the same $\omega\rightarrow 0$ limit get separated out to different values of $\Delta$ in the conformal basis.  At the same time, this is telling us that we need to take seriously the behavior of amplitudes continued off the spectrum that captures standard radiative states.

\begin{figure}[bth!]
\centering
\vspace{-0.5em}
\begin{tikzpicture}[scale=1.5]
\definecolor{darkgreen}{rgb}{.0, 0.5, .1};
\draw[thick](0,2)node[above]{$| \frac{1-\n}{2}, \frac{1-\bar \n}{2}\rangle $} ;
\draw[thick,->](0,2)--node[above left]{$(\bar L_{-1})^{ \bar \n}$} (-1+.05,1+.05);
\draw[thick,->](0,2)--node[above right]{$( L_{-1})^{  \n}$} (2-.05,0+.05);
\draw[thick,->] (-1+.05,1-.05)node[left]{$ (\bar L_{-1})^{\bar \n} | \frac{1-\n}{2}, \frac{1-\bar \n}{2}\rangle $} --node[below left]{$( L_{-1})^{ \n}$} (1-.05,-1+.05) ;
\draw[thick,->] (2-.05,0-.05)node[right]{$ ( L_{-1})^{ \n} | \frac{1-\n}{2}, \frac{1-\bar \n}{2}\rangle $} --node[below right]{$(\bar L_{-1})^{ \bar \n}$} (1.05,-1+.05) ;
\draw[thick](1,-1.2)node[below]{$(L_{-1})^{\n} (\bar L_{-1})^{\bar \n} | \frac{1-\n}{2}, \frac{1-\bar \n}{2}\rangle $};
\filldraw[black] (0,2) circle (2pt) ;
\filldraw[black] (2,0) circle (2pt) ;
\filldraw[black] (-1,1) circle (2pt) ;
\filldraw[black] (1,-1) circle (2pt) ;
\end{tikzpicture}
\caption{Celestial Diamond illustrating the nested submodule structure of $SL(2,\mathbb{C})$ primary descendants that exist for special values of the conformal weights. }
\label{Nested_submodules}
\end{figure}
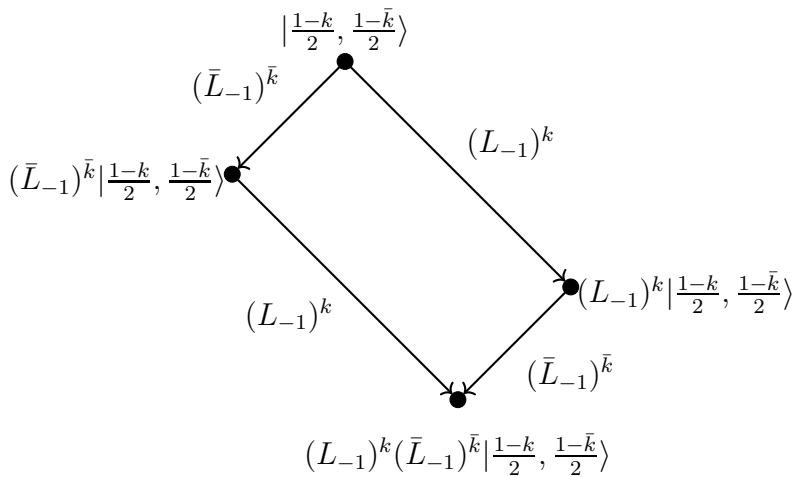

In momentum space, we know there are more soft theorems than the ones with an obvious large gauge symmetry interpretation tabulated in table~\ref{table:Goldstone}.  We note that being conformal primary is effectively a choice of gauge, so while this does not conflict with efforts to attach an asymptotic symmetry to such soft theorems using different gauge choices~\cite{Campiglia:2016jdj,Campiglia:2016hvg,Campiglia:2016efb}, for the purpose of celestial CFT it would be nicer to have a construction that captures these currents in an $SL(2,\mathbb{C})$ covariant manner.  We indeed find such a structure by looking at the $SL(2,\mathbb{C})$ descendants~\cite{Penedones:2015aga,Banerjee:2019aoy,Banerjee:2019tam}. Using the algebra~\eqref{lm} we have
\begin{eqnarray}
    [ L_{1}, (L_{-1})^\n]
     &=\n (L_{-1})^{\n-1} (2L_0+\n-1) \ ,
\end{eqnarray}
so that
\begin{equation}
\label{condition:primary_descendants}
    L_{1} (L_{-1})^\n |h,\bar h \rangle = \n (2h+\n-1) (L_{-1})^{\n-1} |h,\bar
    h \rangle \, .
\end{equation}
We see that a state with weight $h=\frac{1}{2}(1-k)$ for $k\in\mathbb{Z}_>$ will have a primary descendant at level $k$.  An analogous computation holds for the barred modes.  When both the left and right handed weights take on these special half integer values, we get a nested structure of primary descendants as illustrated in figure~\ref{Nested_submodules}.

In the context of celestial CFT these descendancy relations, referred to as celestial diamonds~\cite{Pasterski:2021dqe,Pasterski:2021fjn}, connect soft charges to conformal dressings, capture shadow/helicity redundancies, and account for infinite towers of soft theorems that have appeared in the recent literature~\cite{Guevara:2021abz,Strominger:2021lvk}.  For the purpose of these lectures we will focus on the manner in which they capture the more subleading soft theorems and give generalized currents in CCFT.  Noting that the standard current conservation laws
\be
\p_\bw j_w=0,~~~\p_\bw T_{ww}=0
\ee
as (vanishing) primary descendants that give null state relations in ordinary 2D CFTs, we now have generalized conservation laws that involve higher derivative descendants.   We will see in the next section how to use the symmetries associated to these subleading soft theorems and their descendancy relations to derive recursion relations for celestial OPE coefficients.

Let us now return to the soft theorems we examined in section~\ref{asymptotic} and show how we can use the operators we've defined in this section to reproduce the BMS algebra.   Starting with the conformal primary operators~\eqref{qdelta} identified as asymptotic symmetry currents via table~\ref{table:Goldstone}
\be
P(z)=\lim\limits_{\Delta\rightarrow1}(\Delta-1)\p_{\bz}\mathcal{O}_{\Delta,+2}(z,\bz),~~T(z)= \frac{3!}{2\pi}\int d^2w\frac{1}{(z-w)^4} \lim\limits_{\Delta\rightarrow0}\Delta\mathcal{O}_{\Delta,-2}(w,\bw)
\ee
we define the mode operators
\be\label{poincarelaurent}
P_{a,-1}=\oint dz z^{a+1}P(z),~~~L_{n}=\oint dz z^{n+1}T(z).
\ee
Defining the commutator of of two holomorphic operators via
\be
[A,B](z)=\frac{1}{2\pi i}\oint_z dw A(w)B(z)
\ee
we find that these mode operators obey the BMS algebra~\cite{Fotopoulos:2019vac}
\begin{gather}
  \label{BMSalg}
[L_m,L_n]=(m-n)L_{m+n},~~~[\bar{L}_m,\bar{L}_n]=(m-n)\bar{L}_{m+n},\\
[L_{n},P_{a,b}]=\left(\frac{n-1}{2}-a\right)P_{a+n,b},~~~[\bar{L}_{n},P_{a,b}]=\left(\frac{n-1}{2}-b\right)P_{a,b+n}.
\end{gather}
These statements rely on radial quantization techniques as well as the collinear behavior of amplitudes. In particular, we see the state operator correspondence $|h,\bar h \rangle=\mathcal{O}_{h,\bh}(0,0)|0\rangle$ appearing implicitly in~\eqref{condition:primary_descendants} and symmetry generators defined on contours in~\eqref{BMSalg}. Our understanding of how to apply these tool to celestial CFT is still evolving and we will use section~\ref{literature} to present the current state of affairs. 

A good reference for seeing the details of how to apply these symmetries to explicit amplitudes is~\cite{Banerjee:2020kaa}. Our derivation in the next section will use similar tools, as will the exercise at the end of this section.  Let us take a minute to look at how even just by considering the translation generators we get non-trivial constraints on our celestial amplitudes~\cite{Stieberger:2018onx,Law:2019glh}. Here we will use $\mathcal{A}$ to denote the celestial amplitude arrived at by Mellin transforming the momentum space amplitude $A$.  The constraints coming from Poincar\'e invariance can be written as differential constraints in the $(w_k,\bw_k)$ of the celestial correlators
\be\label{poinconserved}
\mathcal{L}_i\mathcal{A}=\overline{\mathcal{L}}_i\mathcal{A}={\mathcal{P}}_\mu\mathcal{A}=0
\ee
where
\be
\mathcal{L}_i=\sum_k \mathcal{L}_{i,k},~~~\overline{\mathcal{L}}_i=\sum_k \overline{\mathcal{L}}_{i,k},~~~{\mathcal{P}}_\mu=\sum_k {\mathcal{P}}_{\mu,k}
\ee
are the symmetry transformations acting on each of the operators in the celestial correlator. We have already derived the $\mathcal{L}_{i,k}$ in section~\ref{dict}.  Namely we are using this shorthand to denote the right hand side of~\eqref{Liaction} for the $k^{th}$ operator with weights $(h_k,\bar{h}_k)$ at $(w_k,\bw_k)$. We can similarly derive the action of the translation generator $P_\mu$.  Starting from the momentum basis we have
\be
{\mathcal{P}}_{\mu,k}A=\eta_k\omega_k q_{\mu, k}A~~~\Rightarrow~~~ {\mathcal{P}}_{\mu,k}\mathcal{A}=\eta_k q_{\mu,k}e^{\p_{\Delta_k}}\mathcal{A}
\ee
using $p_k=\eta_k \omega_k q_k$ and then noting that the factor of $\omega_k$ shifts the argument of the Mellin transform. While the action of the $\mathcal{L}_i$ is built into the conformal covariance of the celestial amplitude, we get a non-trivial constraint from the $\mathcal{P}_\mu$.  In particular, looking at the $q^+=\frac{1}{2}(q^0+q^3)$ component we have
\be
\sum_k\mathcal{A}(\Delta_1,...,\Delta_k+1,...,\Delta_n)=0.
\ee
For low point Mellin amplitudes translation invariance implies they will have singular support.  We will explore this in more detail in section~\ref{moreccft}.  Let us now turn to a study of collinear limits and how they are captured by celestial OPEs.  Our ability to use symmetries to constrain them will prove to be a key feature of our celestial story.

\subsection{Collinear Limits as OPEs}
We have seen that by tuning the conformal dimensions of our operators we land on celestial currents which obey conformally soft factorization theorems.  We can also consider what happens when we move operators around on the celestial sphere.  From the amplitudes side, having two operators approach each other should correspond to a collinear limit.  In CFT this limit would be governed by an operator product expansion.

We will now show that our holographic dictionary indeed lands us on a celestial OPE that captures the collinear limits of scattering.  Perhaps surprisingly there are two routes one can follow.  The first is to start from the collinear limit of the amplitude and then perform the Mellin transform.  The second route is to use the symmetries we studied in the previous section to establish consistency conditions on the leading OPE coefficients which turn out to have a unique solution.  We will consider both routes in turn.

In keeping with our primary focus on gravitational examples throughout these lectures, we will now look at the collinear limit of two outgoing gravitons.  Aside from the new connection to celestial diamonds, the computation of collinear limits shown here is based off of~\cite{Pate:2019lpp}. See also~\cite{Fotopoulos:2019vac,Raclariu:2021zjz}.  To do so we will need to take a holomorphic collinear limit sending $z_{ij}\rightarrow0$ while holding $\bz_{ij}$ fixed.
From the momentum space amplitude we have
\be
\lim\limits_{z_{ij}\rightarrow0}A_{\ell_1...\ell_n}(p_1,...,p_n)\longrightarrow\sum_{s\in\pm2}\mathrm{Split}^s_{s_is_j}(p_i,p_j)A_{\ell_1...\ell_n}(p_1,...,P,...,p_n)
\ee
where
\be
P^\mu=p_i^\mu+p_j^\mu,~~~\omega_P=\omega_i+\omega_j
\ee
and the collinear splitting factors have the following non-zero components
\be
\mathrm{Split}^2_{22}(p_i,p_j)=-\frac{\kappa}{2}\frac{\bz_{ij}}{z_{ij}}\frac{\omega_P^2}{\omega_i\omega_j},~~~\mathrm{Split}^{-2}_{2-2}(p_i,p_j)=-\frac{\kappa}{2}\frac{\bz_{ij}}{z_{ij}}\frac{\omega_j^3}{\omega_i\omega_P^2}.
\ee
Upon performing the change of variables
\be
\omega_i=t\omega_P,~~~\omega_j=(1-t)\omega_P,
\ee
the Mellin transform hitting the splitting function takes the form
\be\scalemath{0.95}{
\int_0^\infty d\omega_i\omega_i^{\Delta_i-1}\int_0^\infty d\omega_j\omega_j^{\Delta_j-1}\mathrm{Split}^2_{22}\left(\cdot\right)=-\frac{\kappa}{2}\frac{\bz_{ij}}{z_{ij}}\left[\int_0^1 dt t^{\Delta_i-1}(1-t)^{\Delta_j-2}\right]\int_0^\infty d\omega_P \omega_P^{\Delta_i+\Delta_j-1}\left(\cdot\right)}
\ee
and similarly for $\mathrm{Split}^{-2}_{2-2}(p_i,p_j)$.  The term in brackets is the only $t$ dependence and this integral gives the Euler beta function. We thus have
\be\badat{3}\label{ggope}
\mathcal{O}_{\Delta_1,+2}(z_1,\bz_1)\mathcal{O}_{\Delta_1,+2}(z_2,\bz_2)&\sim-\frac{\kappa}{2}\frac{\bz_{12}}{z_{12}}B(\Delta_1-1,\Delta_2-1)\mathcal{O}_{\Delta_1+\Delta_2,+2}(z_2,\bz_2),\\
\mathcal{O}_{\Delta_1,+2}(z_1,\bz_1)\mathcal{O}_{\Delta_1,-2}(z_2,\bz_2)&\sim-\frac{\kappa}{2}\frac{\bz_{12}}{z_{12}}B(\Delta_1-1,\Delta_2+3)\mathcal{O}_{\Delta_1+\Delta_2,-2}(z_2,\bz_2).
\eadat\ee
Let us now see how we can actually derive these 
leading OPE coefficients using celestial symmetries and what we've covered thus far.

First, let us look at how our discussion of the Poincar\'e invariance~\eqref{poinconserved} generalizes.  Looking back at the Ward identity for the subleading soft graviton~\eqref{Tward}, we have an equality between a soft mode insertion which shifts the vacuum and a sum over infinitesimal transformations of the other operators.  Now the soft and hard operators play distinct roles.  From the point of view of celestial diamonds, the soft theorems come from operators at either the left (right) corner.  The right (left) corner is its shadow, and these both descend to the soft charge at the bottom corner.  Looking at the form of the constraint equations~\eqref{constraints}, we see that while the soft charge is a primary descendant it is being equated to a hard charge which should thus also be a primary but is not a descendant.

We can thus think of a Ward identity as living at the bottom corners of the celestial diamond.  If we smear this relation with a function in the kernel of the descendancy relation that takes us from the soft theorem to this soft charge, the contribution from the soft charge will vanish and we are left with a global symmetry of the form
\be
\sum_k\langle \mathcal{O}_1... \delta\mathcal{O}_k...\mathcal{O}_n\rangle=0,~~~\sum_k\langle \mathcal{O}_1... \bar{\delta}\mathcal{O}_k...\mathcal{O}_n\rangle=0
\ee
where we are taking into account that, for all but the leading soft theorems in gauge theory and gravity, there are independent identities coming from each helicity.

The sub-subleading soft graviton appears at $\Delta=-1,J=\pm 2$ and has a level-4 primary descendant.  For the purpose of these lectures, let us start with the corresponding global symmetry transformation identified in~\cite{Pate:2019lpp}
\be
\badat{3}
\delta\mathcal{O}_{\Delta_k,J_k}(z_k,\bz_k)&=-\frac{\kappa}{4}[2h_k(2h_k-1)+8h_kz_k\p_{z_k}+3z_k^2\p_{z_k}^2]\mathcal{O}_{\Delta_k-1,J_k}(z_k,\bz_k),\\
\bar{\delta}\mathcal{O}_{\Delta_k,J_k}(z_k,\bz_k)&=-\frac{\kappa}{4}[2\bar{h}_k(2\bar{h}_k-1)+8\bar{h}_k\bz_k\p_{\bz_k}+3\bz_k^2\p_{\bz_k}^2]\mathcal{O}_{\Delta_k-1,J_k}(z_k,\bz_k).\\
\eadat
\ee
Acting with these symmetries on an ansatz OPE of the form
\be\label{OPEansatz}
\mathcal{O}_{\Delta_1,+2}(z_1,\bz_1)\mathcal{O}_{\Delta_1,\pm2}(z_2,\bz_2)\sim-\frac{\kappa}{2}\frac{\bz_{12}}{z_{12}}E_\pm(\Delta_1,\Delta_2)\mathcal{O}_{\Delta_1+\Delta_2,\pm2}(z_2,\bz_2),
\ee
motivated by dimensional analysis, gives a recursion relation for the $E_\pm(\Delta_1,\Delta_2)$
\be\badat{3}\label{ggreln}
(\Delta_1+1)(\Delta_1-2)E_\pm(\Delta_1-1,\Delta_2)+(\Delta_2\mp2-1)(\Delta_2\mp2)E_\pm(\Delta_1,\Delta_2-1)\\
=(\Delta_1+\Delta_2\mp 2)(\Delta_1+\Delta_2\mp 2-1)E_\pm(\Delta_1,\Delta_2).
\eadat\ee
Combined with another recursion relation from translation invariance that you will derive in the exercises, and the fact that $\Delta\rightarrow 1$ should match the normalization of the leading soft graviton theorem, we land on on~\eqref{ggope}.

We see that symmetries in celestial CFT are quite powerful.  While we have shown how to use the sub-subleading soft graviton to constrain the leading OPE data, the conformal covariance associated to the subleading soft graviton lets us extend this OPE to include descendants.  Following a similar spirit to~\eqref{poincarelaurent} one can investigate the symmetry algebras associated to these OPEs. A common theme in the recent literature is to allow ourselves to complexify the $z_{ij}$ and look at the holomorphic and antiholomorphic algebras separately.  We will highlight some of the recent interesting results that have expanded on this technology in section~\ref{literature}.

\subsection{More Fun with Celestial Amplitudes}\label{moreccft}

In principle, with an understanding of the spectrum and OPE coefficients we have the data of our celestial CFT. The last three subsections have been focused on these aspects of CCFT. The aim is then to be able to use CFT techniques to learn about scattering. So far we have tried to make general statements about features of celestial amplitudes.  In this section, we will take a closer look at the specific form of the transformed amplitudes, focusing on massless $2\rightarrow 2$ scattering.  This will give us a chance to comment on some of the exotic features of celestial CFT and how the can either be tamed or used as a new tool.  We will be following~\cite{Pasterski:2017ylz} and~\cite{Arkani-Hamed:2020gyp}.  Our convention for spinor helicity variables~\eqref{spinorhelicity} matches~\cite{Elvang:2013cua} as in~\cite{Pasterski:2017ylz}.

Recall that the momentum space amplitude contains a distributional factor enforcing momentum conservation
\be
A=M(p_i)\delta^{(4)}(\sum_{i=1}^n p_i).
\ee
This $\delta$-function also undergoes the Mellin transformation.  If we make a change of variables
\be
\omega_i=\mathcal{s}\sigma_i,~~~\mathcal{s}=\sum_i\omega_i,
\ee
we have
\be
\prod_{i=1}^n\int_0^\infty d\omega_i\omega_i^{\Delta_i-1}\left(\cdot\right)=\int_0^\infty d\mathcal{s} \mathcal{s}^{\sum_i\Delta-1}\prod_{i=1}^n\int_0^1\sigma_i^{\Delta_i-1}\delta(\sum_i\sigma_i-1)\left(\cdot\right).
\ee
Because the momentum conserving delta function is homogeneous under simultaneous scale transformations of all the $\omega_i$ we end up with five constraints
\be\label{delta4}
\delta^{(4)}(\sum_i \eta_i \sigma_iq_i)\delta(\sum_i\sigma_i-1)=C(z_i,\bz_i)\prod_{i=1}^{n\le 5}\delta(\sigma_i-\sigma_{i*})
\ee
for some $C(z_i,\bz_i)$ which will impose $5-n$ constraints on the $z_{ij}$.  On the one hand the $n\le 5$ point functions are easy to compute since we have
\be
[ij]=2\sqrt{\omega_i\omega_j}\bz_{ij},~~~\langle ij\rangle=2\eta_i\eta_j\sqrt{\omega_i\omega_j} z_{ij},
\ee
we can just substitute the $\omega_i\rightarrow \mathcal{s}\sigma_{i*}$.  The $\sigma_i$ integrals give indicator functions $\mathbb{1}_{[0,1]}(\sigma_{i*})$ enforcing $\sigma_{i*}\in[0,1]$.  For tree level amplitudes the remaining $\mathcal{s}$ integral just gives a distribution and we are set.

At the same time, however, this also implies the Mellin 2-,3-, and 4- point functions will have singular support.\footnote{Scattering massive particles avoids these kinematical singularities~\cite{Pasterski:2016qvg}.} The two point function is proportional to the inner product for the single particle states, hence a contact term.  To study the 3-point function we need to analytically continue to $\mathbb{R}^{2,2}$, where the amplitude will have $\delta$-function support in either the $\bz_{ij}$ or the $z_{ij}$  for MHV and anti-MHV, respectively~\cite{Pasterski:2017ylz}. These can be tamed with appropriate shadow transforms in $\mathbb{R}^{1,3}$\cite{SS} and light transforms in~$\mathbb{R}^{2,2}$
\cite{ss2,upcomingCP}. 

Here we will focus on the 4-point function.  From global conformal invariance we expect the correlator of four primaries to take the form
\be
\langle \mathcal{O}_{h_1,\bh_1}(z_1,\bz_1)\mathcal{O}_{h_2,\bh_2}(z_2,\bz_2)\mathcal{O}_{h_3,\bh_3}(z_3,\bz_3)\mathcal{O}_{h_4,\bh_4}(z_4,\bz_4)\rangle=\prod\limits_{i<j}z_{ij}^{\frac{h}{3}-h_i-h_j}\bz_{ij}^{\frac{\bh}{3}-\bh_i-\bh_j}f(z,\bz)
\ee
where $h=\sum h_i$, $\bar{h}=\sum \bar{h}_i$ and $z,\bz$ are the cross ratios
\be
z=\frac{z_{12}z_{34}}{z_{13}z_{24}},~~~\bz=\frac{\bz_{12}\bz_{34}}{\bz_{13}\bz_{24}}.
\ee
Meanwhile the generic massless four point amplitude in momentum space takes the form
\be
A=H(z_i,\bz_i)M(s,t)\delta^{(4)}(\sum_{i=1}^4 p_i),~~~H(z_i,\bz_i)=\prod_{i<j}\left(
\eta_i\eta_j\frac{\langle ij\rangle}{[ij]}\right)^{\frac{1}{2}(\frac{1}{3}s-s_i-s_j)}.
\ee
Following the procedure outlined above, the corresponding celestial amplitude reduces to
\be\label{Mmellin}
{\cal A}=X(z_i,\bz_i)\int_0^\infty d\omega \omega^{\beta-1}M(\omega^2,-z^{-1}\omega^2),
\ee
where $\beta=\sum(\Delta_i-1)$ and the kinematical factor $X(z_i,\bz_i)$ is given by
\be
X(z_i,\bz_i)=\prod\limits_{i<j}z_{ij}^{\frac{h}{3}-h_i-h_j}\bz_{ij}^{\frac{\bh}{3}-\bh_i-\bh_j}\theta(z-1)2^{-2-\frac{1}{2}\beta}z^{2-\frac{1}{6}(\beta+4)}(1-z^{-1})^{\frac{1}{6}(\beta+4)}\delta(i(z-\bz)).
\ee
Momentum conservation forces the cross ratio to be real and restricts the range of support for each 4D crossing channel.  We have focused on the $12\rightarrow 34$ channel here.

Celestial amplitudes probe scattering at all energy scales. By writing $\mathcal{A}$ in the form~\eqref{Mmellin} we see how to convert data from an EFT expansion into statements about the analytic structure in the $\Delta\in \mathbb{C}$ plane~\cite{Arkani-Hamed:2020gyp}.  This analyticity is very sensitive to the UV behavior.   As alluded to above, the $\mathcal{s}$ integral for tree amplitudes gives a distribution in $\beta$.  Moreover, it is formally divergent for 4-graviton scattering when the weights are restricted to the principal series spectrum.  However, as pointed out in~\cite{Stieberger:2018edy} adding string form factors saves this. While we have spent most of these lectures focusing on how CCFT lets us take advantage of the symmetries of scattering, this anti-Wilsonian approach hints at yet another way the celestial CFT paradigm might help us identify non-trivial consistency conditions for quantum gravity.

\vspace{1em}

\noindent{\bf Exercise:} 
Celestial Amplitudes are highly constrained by symmetries.  This exercise will provide some practice using the tools introduced in this section.

\vspace{.5em}
\noindent a) Starting from the shadow primaries~\eqref{shadowCPWs}, identify the values of $\Delta$ at which they become pure gauge.  What asymptotic symmetries do they correspond to?  States with these values of $\Delta,J$ will have primary descendants.  Show that the corresponding descendant wavefunctions indeed transform as in~\eqref{Defgenprim}.

\vspace{.5em}

\noindent b) Starting from the ward identity for $\mathcal{P}_\mu$~\eqref{poinconserved} and the OPE ansatz~\eqref{OPEansatz}, derive the recursion relation for $E(\Delta_1,\Delta_2)$ that arises from translation invariance.  Show that
\be
E_\pm(\Delta_1,\Delta_2)=B(\Delta_1-1,\Delta_2\mp 2+1)
\ee
satisfies both this recursion relation and the one in~\eqref{ggreln}.\vspace{.5em}

\noindent c) Consider $2\rightarrow2$ massless scattering. Show that the momentum conserving delta function fixes all four energies to functions of the $\{z_{ij},\bz_{ij}\}$, by solving for the $\sigma_{i*}$ in~\eqref{delta4}.  For what values of the cross ratio does the celestial amplitude have support for each channel $ij\ce{<-->} kl$?

\vspace{.5em}

\noindent d) Consider the low energy expansion for the scattering of photons and gravitons
\be
M(\omega^2,-z^{-1}\omega^2)=\sum_{r\le m;n}a_{n,m,r}(z)\omega^{2n}(G\omega^2)^m\log^r\left(\frac{\omega}{\Lambda_{UV}}\right).
\ee
Using manipulations analogous to our study of soft limits~\eqref{aexp}-\eqref{alim}, identify how these $a_{n,m,r}(z)$ are encoded in the complex $\beta$ plane of the corresponding celestial amplitude.  What limit of scattering does fixed conformal cross ratio probe?

\section{Literature Guide}\label{literature}

The papers~\cite{Strominger:2013lka,Strominger:2013jfa} by Strominger kicked off a series of insights into connections between infrared limits of scattering and asymptotic symmetries of asymptotically flat spacetimes, with~\cite{He:2014laa,He:2014cra} demonstrating an equivalence between soft theorems studied by Weinberg~\cite{Weinberg:1965nx} and the asymptotic analysis of Bondi, van der Burg, Metzner, and Sachs~\cite{Bondi:1962px,Sachs:1962wk,Sachs:1962zza}.  The fact that there were also corresponding physical observables added the third vertex to what became known as the IR triangle~\cite{Strominger:2014pwa,Pasterski:2015tva,Pasterski:2015zua}. Memory effects~\cite{1974SvA....18...17Z,Braginsky:1986ia,gravmem3,Bieri:2013hqa} were the position space incarnation of the soft operators that generated asymptotic symmetry transformations and could serve as experimental probes of the assumptions built into our asymptotic analyses~\cite{Pasterski:2015zua,Susskind:2015hpa,Nichols:2017rqr}. This pattern of relations helped point to missing vertices and natural generalizations~\cite{Lysov:2014csa,Dumitrescu:2015fej,Lysov:2015jrs,Campiglia:2014yka,Campiglia:2015yka,Campiglia:2017dpg,Pate:2017vwa,Ball:2018prg,Himwich:2019dug,Himwich:2019qmj}.  As compared to the leading supertranslation iteration which connected disparate research lines from the 60's, each corner of the superrotation iteration of the triangle was new~\cite{Pasterski:2019ceq}. On the asymptotic symmetry side, there had been various proposals to extend the Lorentz subgroup to include superrotations~\cite{Barnich:2011ct}.  With the viewpoint that soft theorems are equivalent to Ward identities,~\cite{Cachazo:2014fwa} found a new subleading soft theorem for gravity, which was then shown to give a ward identity for superrotations~\cite{Kapec:2014opa}. 
These infinite dimensional symmetry enhancements further led to new insights into old questions about IR divergences and dressings~\cite{Chung:1965zza,Kulish:1970ut,Kapec:2017tkm,Choi:2019rlz}.  Moreover the corresponding symmetries gave black holes a source of soft hair~\cite{Hawking:2015qqa,Hawking:2016msc,Hawking:2016sgy,Strominger:2017aeh}, an idea that has spurred many followup investigations~\cite{Donnay:2015abr,Mirbabayi:2016axw,Donnay:2016ejv,Bousso:2017dny,Haco:2018ske,Donnay:2018ckb,Haco:2019ggi,Rahman:2019bmk,Pasterski:2020xvn}.

A central theme of soft physics story~\cite{Strominger:2017zoo} was the manner in which 4D soft operators looked like 2D currents.  However superrotations were only diagonalized for Rindler energy eigenstates~\cite{Kapec:2014opa}.  The fact that this symmetry provided a candidate stress tensor~\cite{Kapec:2016jld} motivated a change of basis~\cite{Pasterski:2016qvg,Pasterski:2017kqt,Pasterski:2017ylz}.   While, relevant work on the corresponding representations for these scattering states dates back to~\cite{Joos:1962qq}, the Mellin transform that implemented this change of basis was inspired by the work of~\cite{deBoer:2003vf} which considered a holographic reduction of Minkowski space onto hyperbolic slices.  A similar route was followed by~\cite{Cheung:2016iub} which nicely demonstrated features of the IR symmetry currents from this slicing point of view. 

Since then, there has been a concerted effort to understand how features of amplitudes get translated into the celestial basis, as well as how to take advantage of the CFT structure to learn about scattering.  These efforts include: understanding conformally soft modes~\cite{Donnay:2018neh,Ball:2019atb} and their relation~\cite{Donnay:2020guq} to the principal series basis; identifying conformally soft theorems~\cite{Cheung:2016iub,Pate:2019mfs,Adamo:2019ipt,Puhm:2019zbl,Guevara:2019ypd,Fan:2019emx} and celestial OPEs~\cite{Fan:2019emx,Fotopoulos:2019tpe,Pate:2019lpp,Fotopoulos:2019vac,Banerjee:2020kaa,Fan:2020xjj}; extending the construction of conformal primaries to other spins~\cite{Law:2020tsg,Muck:2020wtx,Narayanan:2020amh,Pasterski:2020pdk}; identifying constraints on celestial amplitudes coming from global symmetries~\cite{Stieberger:2018onx,Law:2019glh,Law:2020xcf}, the  conformal multiplet structure~\cite{Banerjee:2019aoy,Banerjee:2019tam,Pasterski:2021dqe,Pasterski:2021fjn}, and 
differential constraints on celestial correlators from soft theorems~\cite{Banerjee:2020kaa,Banerjee:2020zlg}; understanding the celestial analog of double copy relations ~\cite{Casali:2020vuy,Pasterski:2020pdk}, 
 UV behavior~\cite{Stieberger:2018edy,Arkani-Hamed:2020gyp}, loop effects~\cite{Gonzalez:2020tpi,Arkani-Hamed:2020gyp}, and
supersymmetry~\cite{Fotopoulos:2020bqj,Brandhuber:2021nez,Jiang:2021xzy,Hu:2021lrx}; identifying conformal dressings~\cite{Arkani-Hamed:2020gyp,Pasterski:2021dqe}, sources of central extensions~\cite{Nande:2017dba,Himwich:2020rro}, enhanced symmetry algebras~\cite{Guevara:2021abz,Strominger:2021lvk} as well as how they connect to~\cite{Pasterski:2021dqe,Pasterski:2021fjn} effective degrees of freedom for the spontaneously broken symmetries~\cite{Cheung:2016iub,Nguyen:2020hot} and the corresponding vertex operators~\cite{Nande:2017dba,Himwich:2020rro}; defining a state operator correspondence~\cite{Crawley:2021ivb} and investigating conformal block decompositions~\cite{Nandan:2019jas,Fan:2021isc,Atanasov:2021cje}; and more!

\section*{Acknowledgements}
My research is supported by a Sam B. Treiman Fellowship at the Princeton Center for Theoretical Science. These lectures are indebted to the organizers of the 2021 SAGEX PhD School in Amplitudes as well as my fellow celestial holographers and collaborators including: Nima Arkani-Hamed, Alex Atanasov, Adam Ball, Scott Collier, Laura Donnay, Mina Himwich, Daniel Kapec,  Slava Lysov, Zahra Mirzaiyan, Prahar Mitra, Sebastian Mizera, Sruthi Narayanan, Yorgo Pano, Andrea Puhm, Ana-Maria Raclariu, Burkhard Schwab, Shu-Heng Shao, Andy Strominger, Emilio Trevisani, Herman Verlinde, and Sasha Zhiboedov.

\bibliographystyle{utphys}
\bibliography{references}

\end{document}